\documentclass[aps,prd,preprint,eqsecnum,nofootinbib,groupedaddress,showpacs,floatfix,12pt]{revtex4}
\usepackage{graphicx,amssymb,amsbsy,amsfonts,amssymb,amsmath}
\usepackage{hyperref}
\usepackage{multirow}
\usepackage{stackrel}

\newcommand{\be}{\begin{equation}}
\newcommand{\ee}{\end{equation}}

\newcommand{\bea}{\begin{eqnarray}}
\newcommand{\eea}{\end{eqnarray}}

\newcommand{\hm}{\widehat{m}}

\def\Maitre{Ma\^{\i}tre}

\def\sect#1{section~{\ref{#1}}}
\def\sects#1#2{sections~\ref{#1}-\ref{#2}}
\def\eqn#1{eq.~(\ref{#1})}

\def\fig#1{fig.~{\ref{#1}}}

\def\tab#1{table~{\ref{#1}}}

\def\nn{\nonumber}

\def\spa#1.#2{\left\langle#1\,#2\right\rangle}
\def\spb#1.#2{\left[#1\,#2\right]}
\def\spash#1.#2{\spa{\smash{#1}}.{\smash{#2}}}
\def\spbsh#1.#2{\spb{\smash{#1}}.{\smash{#2}}}
\def\sand#1.#2.#3{%
\left\langle\smash{#1}{\vphantom1}^{-}\right|{#2}%
\left|\smash{#3}{\vphantom1}^{-}\right\rangle}
\def\sandpp#1.#2.#3{%
\left\langle\smash{#1}{\vphantom1}^{+}\right|{#2}%
\left|\smash{#3}{\vphantom1}^{+}\right\rangle}
\def\sandpm#1.#2.#3{%
\left\langle\smash{#1}{\vphantom1}^{+}\right|{#2}%
\left|\smash{#3}{\vphantom1}^{-}\right\rangle}
\def\sandmp#1.#2.#3{%
\left\langle\smash{#1}{\vphantom1}^{-}\right|{#2}%
\left|\smash{#3}{\vphantom1}^{+}\right\rangle}

\def\hD{\hat D}
\def\tD{\tilde D}

\def\tN{\tilde N}
\def\hN{\hat N}

\def\tr{{\rm tr}}

\def\trho{\tilde{\rho}}
\def\hrho{\hat{\rho}}
\def\talpha{\tilde{\alpha}}
\def\halpha{\hat{\alpha}}
\def\C{\widehat{C}}
\def\tC{\tilde{C}}
\def\hC{\hat{C}}
\def\tG{\tilde{G}}
\def\tc{\tilde{c}}
\def\hc{\hat{c}}
\def\hOc{\hat{\Omega}_{closed}}
\def\hOe{\hat{\Omega}_{exact}}
\def\Oz{{\Omega}_{zero}}
\def\Oc{\Omega_{closed}}
\def\Oe{\Omega_{exact}}
\def\Om{\Omega_{master}}

\def\te{\tilde{e}}
\def\td{\tilde{d}}
\def\hd{\hat{d}}
\def\tu{\tilde{u}}

\def\ta{\tilde{a}}

\def\tb{\tilde{b}}
\def\tell{\tilde{\ell}}
\def\hell{\hat{\ell}}
\def\tm{\tilde{m}}
\def\tn{\tilde{n}}
\def\tr{\tilde{r}}
\def\tv{\tilde{v}}

\def\tpartial{\tilde{\partial}}
\def\tV{\tilde{V}}

\def\tq{\tilde{q}}
\def\hq{\hat{q}}
\def\tk{\tilde{k}}

\def\tp{\tilde{p}}
\def\hp{\hat{p}}

\def\tl{\tilde{l}}

\def\ax{\alpha}
\def\tax{\tilde{\alpha}}
\def\ti{\tilde{i}}
\def\tj{\tilde{j}}
\def\hj{\hat{j}}
\def\invprops{\rho^0 \cdots \trho^{(\tN-1)}}
\def\invpropsalter{(\rho^0)^{k_0} \cdots (\trho^{(\tN-1)})^{k_{(\tN-1)}}}
\def\invpropslong{ \rho^0 \cdots \rho^{N-1}\, \hrho^0 \cdots \hrho^{(\hN-1)}\,  \trho^0 \cdots \trho^{(\tN-1)}  }
\def\dropinvprops{\rho^0 \cdots \widehat{\rho^i} \cdots  \trho^{(\tN-1)}}
\def\measure{[d\rho,d\alpha]}
\def\cutmeasure{[d\alpha]}
\def\rhomeasure{[d\rho]}
\def\cut{\overset{{\rm cut}}{\longrightarrow}}
\def\ibp{IBP}
\def\Ne{N_{\rm exact}}
\def\Nc{N_{\rm closed}}
\def\Nm{N_{\rm master}}
\def\Nmp{N'_{\rm master}}
\def\Ni{N_{\rm irreducible}}
\def\Ns{N_{\rm ibp-relation}}

\def\Mathematica{{\sc Mathematica}}

\newbox\charbox
\newbox\slabox
\def\s#1{{      
        \setbox\charbox=\hbox{$#1$}
        \setbox\slabox=\hbox{$/$}
        \dimen\charbox=\ht\slabox
        \advance\dimen\charbox by -\dp\slabox
        \advance\dimen\charbox by -\ht\charbox
        \advance\dimen\charbox by \dp\charbox
        \divide\dimen\charbox by 2
        \raise-\dimen\charbox\hbox to \wd\charbox{\hss/\hss}
        \llap{$#1$} }}

\begin{document}

\hbox{\rm\small
FR-PHENO-2015-011
\break}

\title{Two-loop Integrand Decomposition Into Master Integrals
And Surface Terms}

\author{
    Harald Ita
    $\null$
    \\
    {\it Institut of Physics, University of Freiburg\\
    D--79104 Freiburg, Germany}
    } 

\begin{abstract}
Loop amplitudes are conveniently expressed in terms of master integrals whose
coefficients carry the process dependent information.
Similarly before integration, the loop integrands may be expressed as a linear
combination of propagator products with universal numerator-tensors.
Such a decomposition is an important input for the numerical unitarity
approach, which constructs integrand coefficients from on-shell tree
amplitudes.  
We present a new method to organize multi-loop integrands into a
direct sum of terms that integrate to zero (surface terms) and remaining master
integrands. 
This decomposition facilitates a general, numerical unitarity approach for
multi-loop amplitudes circumventing analytic integral reduction.
Vanishing integrals are well known as integration-by-parts identities. Our
construction can be viewed as an explicit construction of a complete set of
integration-by-parts identities excluding doubled propagators. 
Interestingly, a class of `horizontal' identities is singled out which hold as
well for altered propagator powers.
\end{abstract}

\pacs{11.15.Bt, 11.25.Db, 11.55.-m, 12.38.-t, 12.38.Bx \hspace{1cm}}
\maketitle

\numberwithin{equation}{section}

\section{Introduction}
\label{IntroductionSection}
Currently the experiments at the Large Hadron Collider (LHC) are entering a new
energy and luminosity regime. Further upgrades are
expected for a number of years to come. The increasing amount of data will
allow to zoom into known physics and extend the discovery potential for new
physics.  An important ingredient in this quest are precise predictions which
match the measurements' standards. Predictions for key
observables will be necessary, but providing a
larger set of predictions beyond this minimal set will be a clear benefit.  
%
%
Here we present new theoretical methods towards these latter aims.

An important input for precision predictions are first-principle
computations in  perturbative quantum-field theory. In recent years, significant
progress has been made by the theory community in providing predictions through
automated fixed-order computations including quantum
corrections~\cite{VBFNLO,BlackHat,NJet,MadLoop,HelacNLO,GoSam,OpenLoops,Recola}.
These have already lead to a wide class of next-to-leading order (NLO)
predictions for Standard Model processes. In addition, a number of impressive
two-to-two next-to-next-to-leading order (NNLO) results
\cite{2gamNNLO,TopNNLO,VVNNLO,ZgamNNLO,VjetNNLO,HjetNNLO} and further higher-order
predictions \cite{HN3LO} have become available. 
These developments have been driven by a combination of analytic and numerical
advances for computing loop integrals. At one-loop level one can highlight
explicit \cite{DDReduction} and implicit methods
\cite{UnitarityI,GenUnitarityI,GenUnitarityII,GenUnitarityIII,OPP,NumUnitarity,BlackHat}
for reducing (tensor) loop integrals to a standard set of master integrals. 
Similarly, at two-loop level, explicit analytic reduction techniques for
integrals~\cite{IBP,Reduce,Air,Fire,LiteRed,Laporta} play an important role.
Here we discuss methods which bypass analytic integral reduction and
make a numerical approach to multi-loop computations possible.  Such
methods are at the core of the unitarity based approaches
\cite{BlackHat,MZProgram,HelacNLO,NJet,GoSam,MadLoop} to NLO predictions and
allowed to push towards processes with many partons~\cite{W3jB,W3jR,W4j,W5j,5Jets}.  We
are motivated by these results to explore a numerical unitarity approach for
multi-loop amplitudes.

The unitarity method \cite{UnitarityI,
GenUnitarityI,2LoopUnitarityQCD,MultiLoopUnitarity} has continuously provided
cutting edge results for formal as well as phenomenology oriented
multi-loop amplitudes (see e.g. \cite{UVGravity,BDS,SYMUnitarity}  and
\cite{2LoopUnitarityQCD,2LoopAmpl5g}).  This method relates universal
representations of amplitudes in terms of master integrals to full scattering 
amplitudes. By comparing the analytic structure (e.g.  branch cuts) of both
representations the process dependent coefficients of master integrals are
obtained. In this approach, the cutting operation simplifies the loop integrals
to phase-space integrals over on-shell tree-level input. On the one hand, the
strength of this approach arises from efficiently dealing with physical
(on-shell) building blocks. On the other hand, the unitarity approach provides
an implicit integral-reduction mechanism, since by cutting one targets coefficients of
master integrals very directly.

In this article we discuss a numerical variant of the unitarity approach. This
approach is well developed at one-loop level~\cite{OPP,NumUnitarity,BlackHat}
and we extend it to higher-loop orders.  In the numerical approach, the loop
integrations are delayed to the very end of the computation. First one compares
the rational integrands of the Feynman amplitudes with a universal basis of
loop integrands. Delaying the loop integration, however, comes with a price; in
order to maintain the equality of the integrand basis and Feynman amplitudes
additional terms, i.e. surface terms, have to be added to the basis.
These are terms that integrate to zero eventually but are required in
intermediate steps.  The explicit construction of the multi-loop
surface terms is the main result of this article.
%
%
The importance of the surface terms becomes clear in the remaining
computational steps.  Once the coefficients of the integrand basis have been
obtained (through solving linear equations), the loop integration is
performed. In this final step surface terms can be dropped and only the master
integrals have to be provided to yield the loop amplitudes. In this way
the reduction of tensor integrals is accomplished by the integrand
parametrization and is implicit. Given the classification of surface terms we
obtain a general, i.e.  process and multiplicity independent, numerical
algorithm.

Even though the unitarity method operates on-shell, the surface terms have to
be known off-shell. This is required, since already computed unitarity cuts
have to be subtracted in cuts with fewer on-shell propagators in order to avoid
double counting. However, the cut conditions can only be relaxed if we have a
way to take results off-shell.  The prescription to go off-shell is provided
through the integrand parametrization.

%
%

A number of recent developments have advanced the unitarity method to a
promising approach for automated multi-loop computations in QCD.
Parametrizations of loop integrands have been developed recently
\cite{MLoopParamMO,MLoopParamBFZ,MLoopAlgGeo}, which are given in terms of a
minimal basis of irreducible integrands. These parametrizations identify a subset of the
terms (spurious numerators) which integrate to zero, but not all.  Thus, standard
reduction techniques \cite{IBP,Reduce,Air,Fire,Laporta} are required to obtain
a final representation in terms of master integrals.  
Here we put forward a different type of representation of loop
integrands which is organized into surface terms and master integrals.
In our approach, the integral reduction is built-in and does not have to be
performed in a second step, which is important in a numerical approach. 
Furthermore, in analogy to the numerical one-loop unitarity approach, the
integral coefficients can be computed through generalized unitarity cuts
\cite{2LoopMaxCuts} by solving linear equations (e.g. via Fourier transforms).  
This approach does not require integration over multi-dimensional phase spaces
and thus differs from the direct extraction of integral coefficients
\cite{MLoopContours} or possible extensions using the duality between master
integrals and homology cycles of the phase spaces~\cite{YZhangElliptic,MLoopCohom,MLoopContoursCycles}.
Nevertheless, the latter approaches hold the promise to be very efficient once
available in a complete way. 

Technically, a number of new observations lead to the present construction.
First of all, we combine integration-by-parts (IBP) identities
and master integrals to parametrize the loop integrands.  Although this
approach is natural it has not been appreciated for the numerical unitarity
approach so far. 
For the unitarity approach it is best to focus on \ibp{} identities that do not
involve integral topologies with doubled propagators~\cite{KosowerIBP}.  The
identities are obtained from a specialized set of vector fields in
loop-momentum space.  We provide the explicit form of such \ibp{} vectors.
Algorithms to obtain \ibp{} vectors have been suggested in the original
literature and improved in refs.~\cite{sIBP,IBPDiffGeom}.  We give a complete
set of (off-shell) vectors for two-loop topologies.  This construction is
important in order to obtain compact analytic expressions as well as numerical
control throughout momentum space.  The presented construction reproduces the
known results at one-loop level \cite{NumUnitarity}.

Moreover, we find interesting properties of the \ibp{} vectors; using general
coordinate transformations to adjust the integration variables to the integral
topology, \ibp{} vectors can be constructed explicitly with pen and paper.  In
fact, the \ibp{} vectors turn out to come in two types, (complexified)
rotations in momentum space, which leave propagators invariant, and scaling
transformations of the propagators. For our construction the former
`horizontal' generators will be most important.  A special example of such
horizontal vectors was given already in~\cite{KosowerIBP} (based on Gram
determinants).

Although we construct special \ibp{} relations that do not double propagators,
we obtain a much bigger set of \ibp{} relations. In fact, once the numerators
of the horizontal \ibp{} relations are obtained, the propagator powers may be
changed to give new relations. 

We observe that the \ibp{} vectors are tangent vectors to the
unitarity-cut phase spaces. This property allows one to link off-shell 
and on-shell information on unitarity cuts; we show (see
\sect{totdiffpullback}) that surface terms from special \ibp{} relations are as
well surface terms on the unitarity-cut phase spaces.  Interestingly, we
observe as well a Lie-algebra structure which simplifies the construction of
the \ibp{} vectors. This structure appears to be fundamental linking unitarity
cuts to full amplitudes (\sect{intstr}).

Finally, the link between off-shell and on-shell information allows one to relate
master integrands and surface terms to closed and exact holomorphic forms on
the phase spaces, respectively. The number of master integrals is then given
through topological properties of the unitarity-cut phase spaces, that is the
number of closed-modulo-exact forms.
The importance of cohomology for the construction of on-shell \ibp{} identities
has been discussed recently~\cite{MLoopCohom}.  We setup a related but
simplified on-shell approach to multi-dimensional phase spaces.
Although we construct off-shell surface terms, the on-shell perspective
serves as valuable guidance and a cross check. In fact, the on-shell
construction is simpler and its completeness can be verified in a combinatorial
way. We use the on-shell approach to verify that the constructed
surface terms are complete.

Here we focus on planar two-loop integrals, however, our methods can
be extended to non-planar as well as a full $D$-dimensional approach, and we
suggest how the generalizations can be done. In fact, we give \ibp{} vectors
for the non-planar two-loop topologies. Furthermore, we work with generic
    non-vanishing internal and external masses and, thus, capturing much of the
    $D$-dimensional aspects.  We believe that the \ibp{} vectors are sufficient
    as well for most massless integrals; this is plausible assuming that
    factorization limits relate this massive information to the massless
one\footnote{Exceptions may appear when only a single massless leg is attached
to a loop.}.  Finally, we suggest a geometric interpretation of the \ibp{}
vectors (\sect{geointer}) which makes generalizations to multiple loops
natural. 

The article is organized as follows. We start with a heuristic formulation
of the central research question in \sect{Notation}. 
In \sect{LoopMomParam} we present important technical background and methods.
This includes general coordinate transformations of the loop momenta as well as the
discussion of tensor insertions and the
unitarity cuts.  
The off-shell construction of two-loop surface terms is presented in
\sect{offshellconstruction}. The reader interested in the final result should
be able to read starting from \sects{ibpvectorsstandard}{geointer}
which give the \ibp{} vectors, the formula for surface terms and the
vectors' geometric interpretation. In \sect{intstr} we speculate about the
Lie-algebra structure of the \ibp{} vectors. 
In \sect{onshellconstruction} we present the on-shell construction of surface
terms and count master integrands which serves as valuable cross checks of the
off-shell approach.
In \sect{ibponeloop} and \sect{formexample} one-loop examples are given in
order to illustrate the off-shell and on-shell constructions, respectively.
Finally, we close with a summary and a discussion of a number of interesting
future directions.
Technical aspects of differential calculus are discussed in an appendix.

\section{Setup and Notation}
\label{Notation}
We start with an heuristic introduction of the key structures the we will be
dealing with.

\subsection{Loop integrand decomposition} 

\begin{figure}[t]
\includegraphics[clip,scale=0.46]{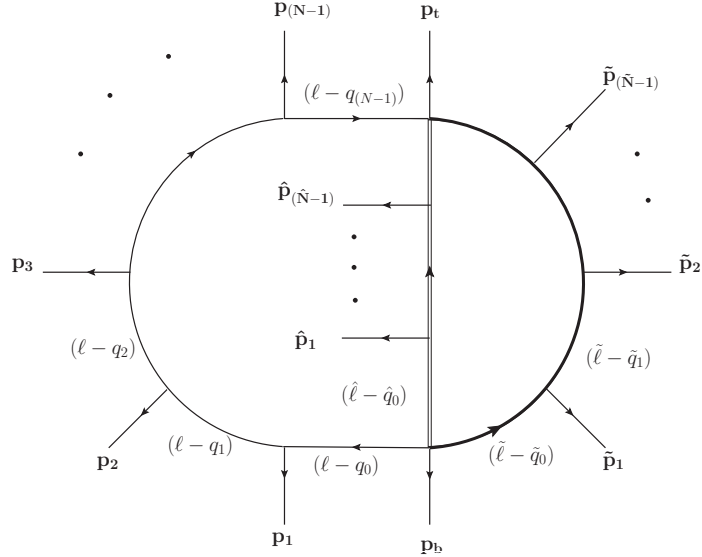}
\caption{
    A generic two-loop integral topology is displayed with the naming
    conventions used in the main text. In order to reuse structures well known
    at one-loop level we interpret the two-loop topology as three rungs. The
    rungs carry loop momentum $\ell$, $\hell$ and $\tell$, and have
    external momenta $p_i$, $\hp_i$ and $\tp_i$ exiting. The rungs are joined
    in two four-point vertices on the top and bottom with external momenta
    $p_t$ and $p_b$ leaving, respectively. 
    Most of our considerations will be focused on the planar case with no
    external momenta $\hp_i$ attached to the central rung at all.  Many
    considerations work analogously for the three rungs and we will often refer
    to the joint variables by dropping 'hat' and 'tilde' super scripts.  }
\label{2loopFigure} \end{figure}

%
We consider two-loop computations with the integral 
topologies as shown in figure \fig{2loopFigure}. Integrals typically include
tensor insertions which are denoted by $t(\ell,\tell)$ giving,
\begin{eqnarray}
    {\cal I}[t]&=&\int d^D\ell d^D\tell\, \frac{ t(\ell,\tell)}{ \invpropslong } \,,\\
    &&\mbox{with}
    \quad \rho^i=(\ell- q_i)^2-m_i^2\,, 
    \quad \hrho^i=(\hell-\hq_i)^2-\hm_i^2\,,
    \quad  \trho^i=(\tell-\tq_i)^2-\tm_i^2\, .\nn
\end{eqnarray}
Momentum conservation is imposed $\hell=-(\ell+\tell-q_0-\hq_0-\tq_0+\tp_b)$.
We will work with dimensional regularization keeping the loop-momentum
dimensions as free parameters. The tensor insertions are assumed to be polynomial in the
loop momenta as it is the case in Feynman amplitudes. 

Given a complete basis of numerator tensors $\{\tm_i(\ell,\tell)\}$ with the
index $i$ labeling the basis elements, one can evaluate tensor integrals by
first decomposing the tensor numerator into the basis,
\begin{eqnarray} t(\ell,\tell)&=&\sum_{ \mbox{\small $i$ $\in$ numerator basis}}
    \td^i\, \tm_i(\ell,\tell) \,, \label{numdecomposition} \end{eqnarray}
with loop-momentum independent coefficients $\td^i$. 
In a second step one has to integrate all the basis tensor
insertions. To this end, typically tensor reduction techniques
\cite{IBP,Reduce,Air,Fire,Laporta} are used to decompose the basis of tensor
integrals into a small set of independent master integrals.

Here we aim to shortcut the step of the tensor reduction by constructing a
particular numerator basis. Following the strategy of one-loop
computations~\cite{OPP,NumUnitarity}, we decompose the numerator tensors into
the tensor insertions associated with master integrals $m_i(\ell,\tell)$ and
surface terms $\hm_j(\ell,\tell)$, which integrate to zero,
\begin{eqnarray}
    t(\ell,\tell)&=&\sum_{ \mbox{\small $i$ $\in$ master integrals}} d^i\,
    m_i(\ell,\tell) + \sum_{ \mbox{\small $j$ $\in$ surface terms}} \hd^j
    \hm_j(\ell,\tell)\,,  \label{tensordecomposition}
\end{eqnarray} 
with the properties,
\begin{eqnarray} {\cal I}_i&:=&\int d^D\ell d^D\tell\,
\frac{m_i(\ell,\tell)}{\invprops}\,,\quad \int d^D\ell d^D\tell \frac{
    \hm_j(\ell,\tell)}{\invprops}=0\,.  \end{eqnarray}
Thus, we directly obtain the decomposition of the initial tensor integral in
terms of master integrals ${\cal I}_i$,
\begin{eqnarray} {\cal I}[t]=\sum_{\mbox{\small $i$ $\in$ master integrals}}
d^i \,{\cal I}_i\,, \end{eqnarray}
while the coefficients $\hd_j$ drop out of the final result.  The surface
    terms contribute only prior to the loop integration expressing for example
    angular correlations. 

Within the (numerical) unitarity approaches one works at the integrand level and
parametrizations (\ref{numdecomposition}) of the loop integrands are required.
Parametrizations have been developed in the recent years
\cite{MLoopParamMO,MLoopParamBFZ,MLoopAlgGeo}, however,
a decomposition in terms of surface terms and master integrals
(\ref{tensordecomposition}) would be important in order to avoid the explicit tensor reduction. 
The construction of the surface terms has so far not been developed
sufficiently and we will provide this missing piece here.

It is important to know the surface terms off-shell, that is all over momentum
space and in particular away from the regions of on-shell propagators. This is
required on the one hand, to ensure that they in fact integrate to zero in the
full loop integrals. 
On the other hand, even in the unitarity approach off-shell information is
required to avoid double counting. That is, given a result for a unitarity cut it
has to be subtracted in cuts with fewer on-shell conditions imposed. (E.g.
at one-loop results from triple cuts have to be subtracted from two-particle
cuts.) This can only be done if the initial cut results can be taken
off-shell in a consistent way. A priori a cut, i.e. a product of on-shell tree
amplitudes,  cannot be taken off-shell. However, once we have used the
cut to compute coefficients of an appropriate loop integrand parametrization
(using on-shell momenta), we can take the latter off shell and subtract it from
daughter cuts.  Steps of this kind are explicit or implicit in almost all
variants of the unitarity approach, but may possibly be circumvented by introducing
phase space integrals or exploiting discrete symmetries \cite{GenUnitarityIII}.
Thus we require an off-shell representation of the loop integrands ideally in
terms of surface terms.
As a final remark we add that the surface terms have to be algebraic
expressions in the loop momenta times propagators. This is the case since they
should represent Feynman amplitudes, which have this property.

Integral relations such as \ibp{} identities have all the properties needed for
surface terms and can be used when available. This fact is very important for a
numerical unitarity approach at higher-loop order.  In principle \ibp{}
relations may be obtained  through standard techniques.  However, for the
numerical unitarity cuts it is beneficial not to consider an integral basis
with doubled propagators. Thus we will construct surface terms from the
specialized \ibp{} relations first introduced in ref.~\cite{KosowerIBP} which
initially do not include integral topologies with doubled propagators. 

An automated construction of the specialized \ibp{} relations has been given
in the original article~\cite{KosowerIBP} and has been advanced in
ref.~\cite{sIBP}. Here we prefer to follow an analytic approach, since we
require additional control over the expressions. That is we do not only need a
compact representation of the surface terms, but we will also need sufficient
numerical control when solving for the integral coefficients in all regions of
phase space.  Nevertheless, it would be instructive to compare the approaches in
detail in the future. A geometric on-shell construction of specialized \ibp{}
relations has been put forward in refs.~\cite{YZhangElliptic,YZhangElliptic2}
which is, however, not suitable for our purpose, since we require the full
off-shell information of the surface terms. Nevertheless, we will use a related
on-shell approach for cross checks below in \sect{onshellconstruction}.

\subsection{Adapted coordinates}
\label{adaptedcoordinates}

Important structures of the loop integrals can be made manifest by using
appropriate integration variables. 
The aim is to change from the loop-momentum components $\{\ell^\mu,\tell^\mu\}$
to using the inverse propagators $\{\rho^i,\trho^j,\hrho^k\}$ as integration
variables~\cite{DOlive}.  Given the missmatch in the number of propagators and loop-momentum
components additional internal variables have to be introduced, which we denote
by indexed $\alpha$'s in the following. The loop integrals are then
given by integrations over the inverse propagators in addition to an internal
integration over the $\alpha$-coordinates,
\begin{eqnarray}
    {\cal I}[t]=\int \frac{\rhomeasure}{\invprops} \,\times\, t(\rho,\alpha)
    \,\mu(\rho,\alpha) \cutmeasure\,,  \label{motivationcohomology}
\end{eqnarray}
where $\mu(\rho,\alpha)$ is the non-trivial integration measure from the
coordinate change. The expression $\rhomeasure$ and  $\cutmeasure$ denote the
differentials of the integration variables, $\rho$'s and $\alpha$'s
respectively.  The insertion $t(\rho,\alpha)$ is the tensor $t(\ell,\tell)$
evaluated in the new coordinates.

It is instructive to consider first the integration over an internal space with the
inverse propagators held fixed. From this perspective we can now use the
properties of the internal space to organize the computation. To give an
example, at one-loop level the internal integration is performed over
spheres~\footnote{Typically these spheres are part of the complex internal
spaces which are tangent bundles of the real spheres $TS_d$.}. We can think
of the function $t(\rho,\alpha)$ being decomposed into a linear combination of
spherical harmonics. Only the constant function gives a non-vanishing integral,
while the higher harmonics integrate to zero. The later can be interpreted as
    surface terms. Thus surface terms are identified  by relating the
    numerator tensors to spherical harmonics.  Fittingly the \ibp{} vectors
    that generate surface terms turn out to be generators of rotations along
    the internal space directions. When acting on an insertion
    $t(\rho,\alpha)$, generators of rotations annihilate the invariant scalar
    parts, and give nontrivial representation of the rotation group otherwise,
    that is, non-trivial spherical harmonics. A similar picture holds at
    higher-loop level but it is also useful to consider the integral in a more
    formal manner. In formal terms we may relate the task of finding
    non-trivial integrals to understanding the cohomology of the internal
    spaces, so that exact forms (total derivatives) in the internal space are
    related to surface terms, while closed but non-exact forms are related to
    the non-vanishing  master integrals. We will find this perspective useful
    when considering generalized cuts of the loop integrals.

Not all vanishing integrals arise from surface terms of the internal space
alone. For example discrete symmetries can lead to further vanishing integrals.
Will not consider the role of discrete symmetries further here, but focus on
the identification of surface terms in the earlier sense.

\subsubsection{Maximal cuts}
Once we transform to adapted coordinates (\ref{motivationcohomology}) as
described above we can naturally make contact with unitarity cuts.  Formally,
unitarity cuts amount to replacing propagators with delta distributions,
$i/\rho\rightarrow \delta(\rho)$.  The insertion of the delta distributions
localizes the integral to vanishing inverse propagators $\rho^k=0$. The
Jacobian factors from the coordinate change to adapted coordinates already
provides the correct measure for the remaining integrations in the internal
variables.  (Details about the coordinate change can be found in
\sect{LoopMomParam}.)

In the maximal cuts of a given integral topology all
independent propagators are formally replaced by delta distributions. Once we
localize to vanishing inverse propagators, the loop momentum takes on-shell
values. Thus the on-shell loop momenta and tensor insertions are obtained by
setting all the $\rho^i$, $\trho^{j}$ and $\hrho^{k}$ to zero. The internal
$\alpha$-variables then are the coordinates of the on-shell loop-momentum
space. We will refer to this subspace as the maximal-cut phase space in the
following. The maximal-cut phase space shares many properties with the surfaces
of fixed propagator values allowing to infer properties of the full loop
integral from on-shell information.

\subsection{Surface terms as specialized \ibp{} identities}
\label{ibpvectors}

Surface terms can be obtained from total derivatives starting from
(sufficiently regular) vector fields $\{u^\mu, \tilde u^\nu\}$,
\begin{eqnarray}\label{surfaceterms}
    \int d^D\ell d^D\tell \left[
    \partial_\mu \left(\frac{ u^\mu\,  t(\ell,\tell) } {\invprops}\right) + 
    \tpartial_\nu \left(\frac{ \tu^\nu \, t(\ell,\tell) } {\invprops}\right)\right] =0\,.
\end{eqnarray}
The components of the vector fields are polynomial in the loop momenta to
obtain relations between Feynman integrals. Typically doubled propagators
appear when the derivatives act on them.

Doubled propagators can be avoided, by a very specific choice of vector-field
insertions \cite{KosowerIBP} fulfilling the equations,
\begin{eqnarray} 
    (u^\mu \partial_{\mu} + \tu^\nu
    \tpartial_{\nu}) \,\rho^i    = f^i(\ell,\tell) \,\rho^i\,, \label{fterms}
\end{eqnarray}
for all inverse propagators $\rho^i$ and similarly for $\trho^j$ and $\hrho^j$
    with independent functions $f^{\tj}(\ell,\tell)$ and
    $f^{\hj}(\ell,\tell)$. Due to the chain rule, the doubled propagators are
    canceled for such special \ibp{} vector fields (\ref{fterms}).  The index
    $i$ on the right-hand side is not summed over.  The functions
    $f^{i}(\ell,\tell)$ are again polynomial in the basic momentum
    contractions.  Typically it is difficult to find this kind of vector
    field, however, we will point out a simplified construction when
    adapted coordinates are used.

The relation \eqn{fterms} has an interesting on-shell interpretation.
When specializing to the on-shell phase spaces with $\rho^i=\trho^j=\hrho^k=0$,
we find that the vector fields $\{u^\mu,\tu^\nu\}$ turn into tangent vectors
along the maximal-cut phase spaces: since the right-hand side of
\eqn{fterms} vanishes, the vectors generate translations that keep the
propagators fixed to zero and thus point along on-shell phase space,
\begin{eqnarray} 
    \{u^\mu,\tu^\nu\} \quad \stackbin[\mbox{\small on-shell}]{}{\xrightarrow{\hspace*{1cm}}
}\quad \mbox{phase-space tangent vector}\,.  
\end{eqnarray}
This property allows one to link off-shell surface terms to on-shell ones as will
be discussed in \sect{totdiffpullback}.
An interpretation of \ibp{} vectors in differential geometry was given as well in
\cite{IBPDiffGeom}.

Although the on-shell perspective is instructive, we eventually need
surface terms that are valid off-shell. To this end we can use the adapted
coordinates of the loop integration. Interestingly, the construction of
specialized \ibp{} vectors can be solved by inspection.  In adapted
coordinates the defining equations are,
\begin{eqnarray}  \label{ibpcondition} 
    && \big( 
    u^{a} \partial_{a} + u^{k}\partial_{k} \big)\,\rho^i= u^{i} =f^i
    \,\rho^i\,,\nn\\ 
    && \Longrightarrow \quad u^i = f^i \,\rho^i \,.  
\end{eqnarray}
which follows from $\partial_k\rho^i=\delta_k^i$. The $i$ labels are not summed
over in the above equation. The notation is explained in more detail in
\sect{planarcuts} and we provide only minimal explanations here.  We use the
shorthand notation that the index $a$ labels the $\alpha$-variables.
Similarly the partial derivative $\partial_k$ denotes either of
$\{\partial_{\rho^i},\partial_{\trho^i},\partial_{\hrho^{i}}\}$.
Furthermore, we suppress function arguments in the $f$-functions;
$f^i:=f^i(\rho,\alpha)$. 

The form of the \ibp{} vectors allows for a natural geometric interpretation:
the $u^a$ components generate to horizontal transformations (with fixed $\rho$)
in the transverse directions and, the $f^i$-components induce local conformal
transformations in the individual propagator directions. 

To summarize, the specialized vector fields have the restriction that the
$\rho^i$-components are proportional to $\rho^i$ and analogously for
tilde/hat-coordinates.  The general form reads,
\begin{eqnarray}  \label{ibpconditionsol} \{u^a,u^i\}=\{ 
        u^a,\rho^i f^i
        \}\,.
\end{eqnarray}
(The doubled labels $i$ are not summed over here.) The component functions
$\{u^a,f^i\}$ are unconstrained apart from the requirement that they are
algebraic in the loop momenta (see also \sect{algebraicdata}).  As we will see
below, given the parametrization of the loop momenta adjusted to the integral
topology, it is straight forward to write down the vector fields and
consequently the surface terms. 

We will often consider vector fields with $f^i=0$ and refer to them as
horizontal \ibp{} vectors. These vectors induce translations along
surfaces of fixed propagator values, which justifies their name.
Horizontal vectors are a natural off-shell continuation of the tangent space of
the unitarity-cut phase spaces,
\begin{eqnarray} 
\mbox{phase-space tangent vector}\,  
\quad \stackbin[\mbox{\small off-shell}]{}{\xrightarrow{\hspace*{1cm}}}
\quad  \{u^a,0\} \,,
\end{eqnarray}
linking on-shell and off-shell data. The conversion between
the two is guided by their Lie-algebra structure (see sections
\ref{ibpvectorsstandard} and \ref{ibpvectorsadapted}).

\section{Loop Momentum Parametrizations}
\label{LoopMomParam}

We will need the explicit form of the general coordinate changes to coordinates
adapted to the integral topologies. 
We will relate the two-loop topology to nested one-loop topologies and reuse
one-loop parametrizations.  The below considerations hold without restrictions
to the dimensionality of the loop momentum which we often suppress.  For simplicity we later
focus on planar integral topologies and assume generic propagator masses and
external masses in order to deal with generic structures rather than special
cases.

\subsection{One-loop topologies}
\label{sect:oneloopparam}

\begin{figure}[t]
\includegraphics[clip,scale=0.46]{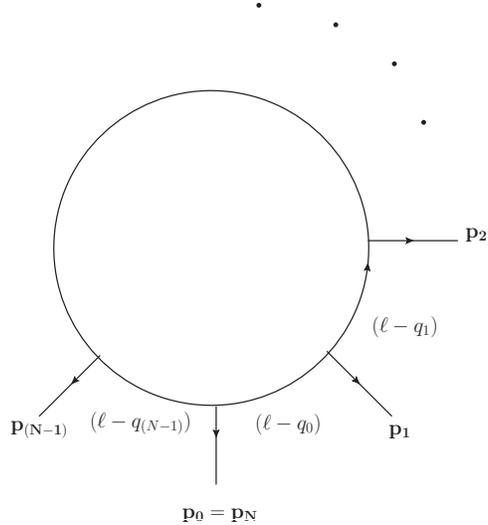}
\caption{
    Conventions for coordinate change for one-loop diagram. Propagator masses
and external momenta enter as parameters.  The loop momentum $\ell$ is parametrized
by the inverse propagators and additional internal variables, in case further
parameters are required. } \label{figoneloop} \end{figure}

We introduce a particular parametrization of the loop momentum adapted to the
topology of the loop diagram in \fig{figoneloop}. The aim is to change
coordinates from the components of the loop momentum $\ell^\mu$ to inverse
propagators $\rho^i$. The construction of such a coordinate transformation can
be obtained from ref.~\cite{NumUnitarity} which we review below.
We use an all-outgoing convention for the external momenta $\{p_{i=1,N}\}$.
The case of generic non-vanishing external and internal masses ($m_i$) is
considered.  Using dimensional regularization in $D$ physical dimensions it
suffices to consider $N$-gon topologies with $N \le D+1$. (This allows loop
momentum dimension to exceed the physical one.) Higher polygons are reducible
using Gram-determinant identities \cite{NVbasis,IntegralsExplicit} and we do
not considered them explicitly.

Inverse propagators will be denoted by $\rho^i$
and are expressed in terms of the loop momentum $\ell^\mu$ by,
\begin{eqnarray}
    \rho^i&=&(\ell-q_i)^2-m_i^2=\left(\ell-\sum_{j=1}^{i} p_j
    \right)^2-m_i^2\,.
\end{eqnarray}
We set the arbitrary constant vector $q_0$ to zero for simplicity.  When fewer
propagators than loop momentum components are present an additional set of
internal (angular) coordinates is required which we denote by $\alpha^i$. The
final result will be the following $D+1$ degrees of freedom,
\begin{eqnarray} 
    \mbox{ inverse propagators:} && \rho^0\,,\dots, \rho^{D_p}\,,\nn\\
    \mbox{ transverse coordinates:} && \alpha^1\,,\alpha^2,\dots,\nn
    \alpha^{D_t}\,,
\end{eqnarray}
with $D_p=N-1$ and $D_p+D_t=D$ and one additional quadratic constraint.  The
form of the quadratic equation will be discussed below and is given in
\eqn{eqn:quadoneloop}.

Explicitly, the coordinate change to adapted coordinates is given by,
\begin{eqnarray}
\label{oneloopparam}
&&    \ell=V(\rho)+\sum_{a=1}^{D_t} n_a\alpha^a\,,\quad V(\rho):=\sum_{i=1}^{D_p} r_i\, v^i\,, \nonumber\\
&& \quad\quad   r_i:= -\frac{1}{2}\left( (\rho^i+m_i^2-q_i^2)-(\rho^{i-1}+m_{i-1}^2-q_{i-1}^2)\right)\,,
\quad q_i= \sum_{j=1}^i p_j\,.\label{coordinatetrafo}
\end{eqnarray}
Here the vectors $v^i$ and $n_a$ are elements of the van Neerven-Vermaseren (NV)
basis~\cite{NVbasis} which we introduce momentarily. This basis is adapted to
the integral topology and splits momentum space into a $D_p=(N-1)$
dimensional 'physical' space spanned by the external momenta and a
$D_t=(D-D_p)$ dimensional 'transverse' space. 
For the considered $N$-gons we have $D_p$ linear independent external momenta
in the set $\{p_{i=1,..,(D_p+1)}\}$ due to momentum conservation. (We will also
use the convenient notation $p_0=p_N$.) 
Using the inverse of the $D_p$-dimensional Gram matrix $G_{ij}=(p_i,p_j)$ the
vectors dual to the external momenta are introduced with $v^i=(G^{-1})^{ij}
p_j$ so that they fulfill $(v^i,p_j)=\delta^i_j$.  In
transverse space an orthonormal basis is used $\{n_{a=1,..,D_t}\}\,,$ with
$(n_a,p_j)=0$ and $(n_a,n_b)=\delta_{ab}$. The transverse basis is not unique
and may be changed by (complex) orthogonal rotations.  The linear dependence of
the vectors $\{p_i\}$ and $\{v^j\}$ implies as well $(v^i,n_a)=0$.
Explicit analytic expressions for the NV basis may also be found in
ref.~\cite{NumUnitarity}.  Particularly compact expressions for the basis
decomposition can be obtained in spinor-helicity notation and is inherent in
most literature considering analytic unitarity methods (see
e.g.~\cite{GenUnitarityIII,BlackHat}).

The parametrization (\ref{oneloopparam}) solves the linear equations,
\begin{eqnarray} \label{rdefs}
(p_i,\ell) = -\frac{1}{2}\left(
    (\rho^i+m_i^2-q_i^2)-(\rho^{i-1}+m_{i-1}^2-q_{i-1}^2)\right) = r_i \,,
\end{eqnarray}
due to the vectors $p_i$ and $v^j$ being dual. An additional quadratic
constraint equation for the internal variables $\alpha^a$ is imposed to make
sure that the square of the loop momentum gives the inverse propagator,
\begin{eqnarray} \label{eqn:quadoneloop} 
    c(\rho,\alpha) &=&  (\ell^2-m_0^2) - \rho^0 = \sum_{a=1}^{D_t} 
        (\alpha^a)^2  + C(\rho)=0\,,\\
        &&C(\rho)=V(\rho)^2-\rho^0-m_0^2\,,\nn
\end{eqnarray}
with $V(\rho)^2 = (G^{-1})^{ij} r_i r_j$. With the linear and quadratic
equations fulfilled this parametrization returns the correct values for all
inverse propagators.  Most of the physical information is condensed neatly into
the quadratic equation through the Gram matrix as well as its momentum and mass
dependence. 

A few remarks can be added here: The equation $c(\rho,\alpha)=0$ allows one to
eliminate one  $\alpha^a$ in terms of the inverse propagators and the remaining
transverse coordinates.  Explicit solutions can be obtained using for example
light cone coordinates,
\begin{eqnarray}
    &&\alpha^{1}=\frac{1}{2} \left(t - \frac{C(\rho) + \sum_{a=3}^{D_t}
    (\alpha^a)^2}{t}\right) \quad \alpha^{2}=\frac{i}{2} \left(t +
    \frac{C(\rho) + \sum_{a=3}^{D_t} (\alpha^a)^2}{t}\right)\,, \nn
\end{eqnarray}
for $D_t>1$, and the sum-term is dropped for $D_t=2$. Here both $\alpha^1$ and
    $\alpha^2$ were traded for a new complex coordinate $t$.  For $D_t=1$ one
    can solve for $\alpha^1$ directly to obtain, $$\alpha^1= \pm i
    \sqrt{C(\rho)}\,,$$ where the internal manifold degenerates to two distinct
    points, i.e. a zero-dimensional sphere.  However, it is often useful to
    consider the loop momentum as a hyper surface in $\{\alpha^a, \rho^i\}$
    space, without using an explicit solution of the quadratic equation
    inserted.

The inverse coordinate change is given by, 
\begin{eqnarray}
    \rho^i(\ell)&=&(\ell-q_i)^2-m_i^2\,, \quad \alpha^a(\ell)=(n^a,\ell)\,.
\end{eqnarray}

For the above loop-momentum parametrization the maximal-cut on-shell conditions
are implemented by setting the inverse propagators to zero, $\rho^i\rightarrow
0$.

\subsection{Two-loop topologies}
\label{planarcuts}

Loop-momentum parametrizations can be obtained by decomposing
multi-loop diagrams into sub diagrams which admit one-loop
parametrizations. To be specific, any rung in a multi-loop diagram admits a
one-loop coordinate transformation yielding sets of internal coordinates and
quadratic equations.  When the rungs are joined in vertices the
momentum-conservation conditions impose additional linear equations adding sets
of linear equations. Planar as well as non-planar multi-loop parametrizations
may be obtained in this way.  We will focus first on the planar diagrams. 

The generic two-loop topology is displayed in \fig{2loopFigure}. The planar
integrals are obtained by specializing to the case where no external momenta
are attached to the central rung, i.e. $\hN=0$.  In an approach best adapted to
planar diagrams we consider the left and right one-loop sub diagrams in the
figure and ignore the central rung at first. For the left loop the following
external momenta and propagators are used, 
\begin{eqnarray}
    &&  \{p_1\,,\dots , p_{N-1}, p_N=-(p_1 +\dots + p_{(N-1)})\}\,, \nn\\ &&
    \rho^i=(\ell-q_i)^2-m_i^2\,,\quad i=1,\dots N\,. \nn
\end{eqnarray}
The quantities for the one-loop coordinate transformation $r_i$, $v^i$,
$G_{ij}=(p_i,p_j)$ and $n_a$ are obtained as in \sect{sect:oneloopparam}.
Analogously, for the right loop we apply the one-loop transformation with the
following list of external momenta inserted,
\begin{eqnarray} \label{rungmap}
    &&  \{\tp_1\,,\dots , \tp_{\tN-1}, \tp_N=-(\tp_1 +\dots +
    \tp_{(\tN-1)})\}\,, \\ &&  \trho^i=(\tell-\tq_i)^2-\tm_i^2\,,\quad
    i=1,\dots \tN\,.\nn
\end{eqnarray}
Now we denote the parameters and functions by $\tr_i$, $\tv^i$,
$\tG_{ij}=(\tp_i,\tp_j)$ and $\tn_{a}$. It is often convenient to distinguish
the vectors and derived quantities by their index only, e.g.
$\alpha_{\ta}=\talpha_{\ta}$.

With these transformations we have the loop momenta parametrized in terms of
$\{D_p=(N-1),\tD_p=(\tN-1)\}$ inverse propagators $\{\rho^i,\trho^j\}$ with
$1\le i \le D_p$ and $1\le j\le \tD_p$. In addition $D_t=(D-D_p)$ and $\tD_t=(D-\tD_p)$
internal coordinates $\{\alpha^a\}$ and $\{\talpha^{\ta}\}$ are introduced, respectively. 

Explicitly the loop momenta are given by,
\begin{eqnarray} \label{2loopmoms} 
    \ell&=& V(\rho)+\sum_{a=1}^{D_t}\alpha^a \,n_a\,, \quad \tilde 
    \ell=\tV(\trho)+ \sum_{\ta=1}^{\tilde D_t}\tilde\alpha^{\ta} \,
    \tn_{\ta}\,,
\end{eqnarray} 
with  $V(\rho)=\sum_{i=1}^{D_p} r^i v_i$ and $\tV(\trho)=\sum_{i=1}^{\tD_p}
\tr^i \tv_i$. The internal coordinates have to fulfill the conditions,
\begin{eqnarray} \label{symequations}
    c(\rho,\alpha) = \sum_{a=1}^{D_t} (\alpha^a)^2 + C(\rho) = 0 \,, \quad 
    \tc(\trho,\talpha) = \sum_{\ta=1}^{\tD_t} (\talpha^{\ta})^2 + \tC(\trho) =
    0 \,,
\end{eqnarray} 
where $C(\rho)=(V(\rho)^2 - \rho^0-m_0^2)),$ and
$\tC(\trho)=(\tV(\trho)^2 - \trho^0-\tm_0^2)$ setting $q_0=\tq_0=0$. 

There is one remaining transformation required; in order to express one
internal degree of freedom from $\{\alpha^a,\talpha^{\ta}\}$ in terms of the
inverse propagator $\hrho^0$ of the central rung we have the relation,
\begin{eqnarray} \label{rungrelation}
    \hc(\rho,\alpha,\trho,\talpha) &=&  \left(\ell+\tilde\ell+ p_b\right)^2 -    \hrho^0 - \hm_0^2 
    \,\nn \\ &=&          2\,(\ell
        + p_b, \tilde \ell + p_b) - \nn \\ && -\hrho^0 -\hm_0^2  +
    \rho^{0} + m_{0}^2  + \trho^{0} + \tm_{0}^2 - p_b^2\,.
\end{eqnarray}
It will be helpful to make the dependence on the $\alpha$-coordinates more
explicit, by inserting the form of the loop momenta. We obtain the quadratic
equation, \begin{eqnarray} \label{cond:rung}
    \hc = \alpha^a \talpha^{\ta} \, \C_{a\ta}+ \alpha^{a}\, \C_a +
    \talpha^{\ta} \, \C_{\ta} + \C\,,
\end{eqnarray}
with the definitions,
\begin{eqnarray} \label{genformrung}
&& \C_{a}^{\tk} = 2 (n_a,\tv^{\tk})\,, \quad \C^{k}_{\ta} = 2
(v^k,\tn_{\ta})\,, \quad \C_{a \ta } = 2 (n_a,\tn_{\ta})\,, \quad \C^{k \tk }
= 2 (v^k,\tv^{\tk})\,, \nn\\ 
&& \C_a = \C_{a\ta}\, \talpha_0^{\ta} +
\C_{a}^{\tk}\,(\tr +\tr_0)_{\tk} \,,\quad \C_{\ta} = \C^T_{\ta a}\,\alpha_0^{a}
+ \C_{\ta}^{k} \, (r + r_{0})_k \,,\nn\\
&& \alpha_0^a= (p_b,n_a)\,,\quad \talpha_0^{\ta}= (p_b,\tn_{\ta})\,, \quad
r_{0,l}= (p_b,p_l)\,,\quad \tr_{0,\tl}= (p_b,\tp_{\tl})\,, \\ 
&& \C = -\hrho^0 - \hm_0^2 + \rho^0 + m_0^2  +\trho^0 + \tm_0^2 + p_b^2 +
(r_k \tr _{\tk} + r_{0,k} \tr_{\tk} + r_k \tr_{0,\tk}) \C^{k\tk} \,,\nn
\end{eqnarray}
where $\C_{a},\C_{\ta}$ and $\C$ depend on external kinematics and inverse
propagators, while the two-index terms $\C_{a\ta}, \C_{a}^{\tk}, \C^{k}_{\ta}$
and $\C^{k \tk }$ depend only on the external momenta. The latter matrices
quantify the alignment of the physical and transverse spaces of the respective
one-loop sub diagrams. 

In general, complex orthogonal transformations may be used to rotate the basis vectors 
of the internal spaces (acting on $a$-labels) and transform the above constraint
(\ref{genformrung}) to canonical form.  We will discuss the relation of
integral topologies and the form of these equations in more detail below. 

For some  topologies it is possible to align the basis for the transverse
spaces of left and right loop in \fig{2loopFigure}. This leads to a block diagonal form of $C_{a \ta
}$ and vanishing entries in $C_{k \tk }$ and  $C_{a\tk}$. Similarly the
($D>4$)-dimensional components of the transverse space can be aligned. Rotation
symmetries in these independent parts of transverse space are then manifest and
simplify the quadratic equations.

In summary, we have traded the loop momenta $\ell^\mu$ and
$\tell^\nu$ for the following coordinates and conditions,
\begin{eqnarray} 
    \mbox{ inverse propagators:} && \rho^0\,,\dots, \rho^{D_p}\,,
    \trho^0\,,...\,,\rho^{\tD_p}\quad \mbox{and}\quad \hrho^0  \,,\nn\\
    \mbox{ transverse coordinates:} && \alpha^1\,,\dots, \alpha^{D_t}\,,
    \talpha^1\,,\dots, \talpha^{\tD_t}\,,\nn\\ 
    \mbox{ quadratic equations:} && c=0\,,\quad \tc=0\quad\mbox{and} \quad
    \hc=0 \,.\label{eqn:quadtwoloop}
\end{eqnarray}
The quadratic equations $c=\tc=\hc=0$ have to be solved for the internal
coordinates $\alpha$. Instead of finding explicit solutions it is often useful
to think of the loop-momentum space as the sub manifolds defined by the
quadratic equations in the unconstrained coordinate space of the  $\rho$'s and
$\alpha$'s.

\subsection{Non-planar parametrization}
\label{nonplanparam}

Two equivalent ways to consider  non-planar topologies will be discussed. The
first emphasizes the general structure of multi-loop parametrizations, the second is
most convenient for two-loop topologies being an adaptation of the planar
setup.
\begin{enumerate}
\item Generic parametrization: The non-planar two-loop topology can be viewed
as multiple rungs which are joined in vertices; for our notation we refer to
\fig{2loopFigure} (see also later in \fig{multirungFig}). The individual rungs
carry the loop momenta $\ell,\tell$ and $\hell$, respectively, which are
related by momentum conservation.  Each rung can be parametrized using one-loop
parametrizations to give three sets of $\alpha$-coordinates and
$\rho$-coordinates constrained by three quadratic equations. Compared to the
planar case we obtain as well,
\begin{eqnarray}
    \hc(\hrho,\halpha) = \sum_{a=1}^{\hD_t} (\halpha^a)^2 + \hC(\hrho) = 0 \,,
    \quad 
\end{eqnarray}
with all functions being natural generalizations of the ones above
(\ref{2loopmoms}) with `tildes' replaced by `hats'.  In a second step momentum
conservation,
\begin{eqnarray}
(\ell-q_0) + (\tell-\tq_0) + (\hell-\hq_0) + p_b = 0\,, 
\end{eqnarray}
is imposed to relate the transverse coordinates of the individual rungs. In
this way one obtains additional coordinates and equations, while the concepts
remain the same. Considering multi-loop topologies amounts to adding further
rungs and vertices in a similar way.

\item Planar induced parametrization: alternatively we can start from a planar
parametrization of the loop momenta and include additional relations to
transform the transverse coordinates ($\alpha$-coordinates) to inverse
propagators.

As far as loop momentum parametrizations are concerned rungs can be exchanged,
so we can always consider the central rung to have the least amount of external
momenta attached; $\hN \le N$ and $\hN \le \tN$. Given that we can have at most
eight propagators we have $\hN \le 2$. Thus, compared to the planar case only
one additional inverse propagator variable is required.
The constrains from the central rung are explicitly given by,
\begin{eqnarray}
\hc(\ell,\tell)&=& (\hell-\hq_0)^2 - \hm_0^2 -\hrho^0   = (\ell+\tell-q_0-\tq_0
+ p_b)^2 -(\hrho^0 + \hm_0^2) \,,\nn\\
\hc''(\ell,\tell) &=& (\hell-\hq_1)^2 - \hm_1^2 -\hrho^1=
 (\ell+\tell-q_0-\tq_0 + p_b +\hp_1 )^2 -(\hrho^1 + \hm_1^2) \nn\\
&=& \hc(\ell,\tell) + (\rho^0 + \tm_0^2) - (\rho^1 + \hm_1^2) + 2 (
\ell+\tell-q_0-\tq_0 + p_b ,\hp_1 )\,.\nn
\end{eqnarray}
While the first constraint is the one already present in the planar topologies,
the second one gives one additional linear equation for the loop momenta. It is
useful to introduce the simplified constraint explicitly,
\begin{eqnarray} \label{nonplanarrel}
\hc'(\ell,\tell) &=& 
(\rho^0 + \tm_0^2) - (\rho^1 + \hm_1^2) +  2 ( \ell+\tell-q_0-\tq_0 + p_b
,\hp_1 )\,.
\end{eqnarray}
\end{enumerate} 
We will return to the non-planar cases when discussing explicit \ibp{} vectors
in sections \ref{ibpvectorsadapted} and \ref{ibpvectorsstandard}.

\subsection{A useful integral classification}
\label{intclass}

It will be useful to refer to individual integral topologies. In principle
it is sufficient to specify the number of rung momenta $(N-1,\tN-1,\hN-1)$, with
conventions as in \fig{2loopFigure} and stating which of the vectors $p_t$ and $p_b$
vanish. We use the following terminology for the topologies,
\begin{eqnarray}
 \mbox{generic:} &&p_t\ne 0\,,\quad p_b\ne 0  \nn\\
 \mbox{semi-generic:} && p_t=0\,,\quad p_b\ne 0 \,,\quad \mbox{or}\quad
 p_t\neq0\,,\quad p_b= 0  \nn\\ 
 \mbox{simple:} &&p_t=0\,,\quad p_b=0\,.\nn
\end{eqnarray}
These sub classes differ in the number of dependent external momenta which are
present in the set $\{p_i,\tp_i,\hp_i\}$.

For planar topologies we label the integral topologies by only two numbers
which specify external legs $(n,\tn)=(N-1,\tN-1)$.

If the number of linear independent external momenta is smaller than the
physical dimension the transverse spaces of left and right loops overlap and a
common transverse space can be defined. We will assume that the transverse
NV-vectors are aligned whenever possible. 

\subsection{Algebraic Data}
\label{algebraicdata}

Tensor insertions from Feynman rules give algebraic functions that can be
obtained by contracting loop momenta with themselves or with tensors derived
from external momenta.  These terms are natural in canonical coordinates in
momentum space.  When using the adapted coordinates we have to make sure that
we deal with expressions that arise from coordinate transformations of such
algebraic functions.  

It turns out that tensor insertions are in fact in one-to-one correspondence
with polynomial expressions in the adapted coordinates. 
This can be shown as follows: On the one hand, adapted coordinates are
conventional loop momentum contractions being inverse propagators $\rho^i$ or
contractions of the form $\alpha^a=(n_a,\ell)$.  Consequently polynomials in
these coordinates are polynomial in the loop momenta.  On the other hand, these
functions are sufficient to represent all loop momentum contractions: given an
expression $(t_{\mu_1...  \nu_k}\ell^{\mu_1} ... \,\tell^{\nu_k})$ we can
insert the completeness relations $(\delta_\mu^\nu=n^a_\mu n^{a\nu}+v^k_\mu
p_k^\nu)$ and $(\delta_\mu^\nu=\tn^{\ta}_\mu \tn^{\ta\nu}+\tv^k_\mu \tp_k^\nu)$
into the contractions with $\ell^\mu$ and $\tell^\nu$, respectively.  The
resulting terms $(p_i,\ell)$ and $(n_a,\ell)$ (and similar for the
tilde-coordinates) give $r^i$ (\ref{rdefs}) and $\alpha^a$ respectively.
Both are polynomial in $\rho$'s and $\alpha$'s, so that any tensor can be
expressed in terms of these coordinates.

Thus we can trade any tensor in canonical coordinates for polynomials written in terms
of inverse propagators and contractions with transverse vectors,
\begin{eqnarray}\label{tensorbasis}
(t_{\mu_1...  \nu_k}\ell^{\mu_1} ... \,\tell^{\nu_k}) \quad
\longleftrightarrow\quad 
\prod_{a,\ta,l,\tl} (\alpha^a)^{k_a}
\,(\talpha^{\ta})^{k_{\ta}}\times (\rho^l)^{k_l}(\trho^{\tl})^{k_{\tl}}  \,.
\end{eqnarray}
One may as well include $\hrho$ and $\halpha$ variables, which however can be
converted to the above monomials. The variables $k_i$ denote non-negative integers.

In order to obtain algebraic surface terms through \ibp{}
identities one has to consider algebraic vector fields in momentum space.  Such
vectors $\{u^\mu,\tu^\nu\}$ are defined to yield algebraic functions
$t'(\ell,\tell)$ upon taking directional derivatives of algebraic functions
$t(\ell,\tell)$, 
\begin{equation} 
(u^\mu\partial_\mu+\tu^\nu \tpartial_\nu)\, t(\ell,\tell) = t'(\ell,\tell)\,.
\end{equation}
We may use general coordinate transformations to obtain vector fields in adapted
coordinates, however, it is preferable to construct them directly using the above
definition in adapted coordinates,
\begin{equation} 
(u^a\partial_a+u^k\partial_k)\, t(\alpha,\rho) = t'(\alpha,\rho)\,.
\end{equation}
(We use the shorthand notation for partial derivatives suppressing 'hats' and
'tildes' as in \eqn{ibpconditionsol}.) One can show that the
components $u^a$ and
$u^{k}$ of algebraic vector fields have to be algebraic functions by acting
one-by-one on the $\alpha$ and $\rho$ coordinates.  There is a further condition: above we
worked in the coordinate space prior to imposing the conditions $c=\tc=\hc=0$.
Consistent vectors have to be tangent vectors to this surface, which defines
the physical momentum space.  This means we have to impose the three equations,
\begin{equation} \label{algebraiccondition}
(u^a\partial_a+u^k\partial_k)\, \{c,\tc,\hc\} = 0\,,
\end{equation}
to obtain an algebraic vector field. In the
non-planar case we have to include one further analogous relation for $\hc'$
(\ref{nonplanarrel}).

From these definitions it is clear that multiplying an algebraic vector field
by an algebraic function yields again an algebraic vector field. 

Finally, given that we deal with integration, we will use differential forms
and outer derivatives. As usually, these are defined as linear functions that
return the components of vector fields. The differentials,
$$\{d\alpha^a,d\rho^k\}$$ are algebraic, yielding algebraic functions when
acting as linear forms on algebraic vector fields with
$d\alpha^a(\partial_b)=\delta^a_b$, $d\rho^k(\partial_l)=\delta^k_l$ etc. We
use again a single label which runs as well over hat and tilde variables.

The 1-forms are not independent due to the relations, 
\begin{equation} \label{onshellrelationforms}
dc=\frac{\partial c}{\partial \alpha^a}d\alpha^a + \frac{\partial
c}{\partial \rho^k}d\rho^k  =0\,,
\end{equation}
and analogously for $\tc$ and the $\hc$ relations.  Wedge products can be used
to generate the full set of differential forms in adapted coordinates.

\subsection{Function ring and  numerator tensors}
\label{numbasis}

We will require a complete set of tensor insertions for a given integral
topology limited only by power counting (of typical field theories). Systematic
constructions of such a basis of tensor insertions can be found at one-loop
level in in ref.~\cite{NumUnitarity} (see also \cite{RevUnitarity}) and for
multi-loop topologies in refs.~\cite{MLoopParamBFZ, MLoopAlgGeo, MLoopParamMO}.
We will review the construction and adjust the notation to our setup. The use
of the adapted coordinates makes the construction of irreducible tensor
insertions very direct, so that it can often be obtained by hand.

For a given integral topology not all tensor insertions are viewed as
independent; inserting an inverse propagator allows one to cancel a propagator and
leads to a reduced topology. Thus we can consider numerator tensors modulo
inverse propagators. This implies that for the construction of independent
numerators inverse propagators are best treated as equivalent to zero
$\rho^k\sim0$. Comparing to (\ref{tensorbasis}) we can
proceed in two steps. First, we identify an over complete list of numerator
tensors as the  polynomials in the $\alpha$ coordinates
\begin{eqnarray}\label{eqn:numbasis}
\prod_{a,\ta} (\alpha^a)^{k_a} \,(\talpha^{\ta})^{k_{\ta}} = 
\prod_{a,\ta} (n_a,\ell)^{k_a} \,(\tn_{\ta},\tell)^{k_{\ta}}\,,
\end{eqnarray}
which are written as well in tensor notation for convenience.
The integers $k_a$ and $k_{\ta}$ take positive values limited above by
power counting. In a second step we use the relations $(c=\tc=\hc=0)$, which
allow to turn certain polynomials into inverse propagators or monomials of
lower degree, thus reducing the independent tensors insertion further. 

It is important to note that these equivalence relations amount to imposing the
on-shell conditions, with all inverse propagators set to zero.  Thus, as far as
the construction of a basis of numerator tensors is concerned, linear
independent numerator tensors remain independent
functions when considered on-shell on the maximal-cut phase spaces.

However, not all questions can be answered modulo lower topologies and
off-shell information is important. For example, considering the tensor
insertion of an inverse propagator, we have an 'uninteresting' tensor insertion,
\begin{eqnarray}
t(\ell,\tell)=\rho^k \sim 0
\end{eqnarray}
and even obtain zero on the maximal cut.  However, considering the loop
integral, we clearly obtain a scalar integral of lower topology yielding a
non-vanishing result,
\begin{eqnarray}
\int d^D\ell d^D\tell \frac{\rho^k}{\invprops} = \int d^D\ell d^D\tell
\frac{1}{\rho^1 \cdots \widehat{\rho^k} \cdots \trho^{(\tN-1)} } \neq 0\,,
\end{eqnarray}
with  $\widehat{\rho^k}$ denoting the omission of the inverse propagator in the
numerator.  Thus when surface terms are analyzed, we have to work
off-shell, although we may obtain guidance from related on-shell
questions. 

\subsection{Total derivatives and master-integral count}
\label{totdiffpullback}

Here we discuss the relation between computing the total derivatives and
cutting all propagators of an integrand. Cutting propagators amounts to
replacing the propagators with delta-distributions ($i/\rho\rightarrow
\delta(\rho)$). When a tensor integral is written in adapted coordinates (as
done in \eqn{motivationcohomology}) the operation of cutting omits all
propagators as well as the integration measure $[d\rho]$ and sets the inverse
propagators to zero,
\begin{eqnarray}
    \int 
    \frac{\rhomeasure}{\invprops} \,\times\,
    t(\rho,\alpha)\,\mu(\rho,\alpha)\cutmeasure   \quad \cut\quad \int
    t(0,\alpha) \,\mu(0,\alpha)   \cutmeasure\,.
\end{eqnarray}
The function $\mu(\rho,\alpha)$ is a measure factor arising from transforming
canonical coordinates $\ell^\mu$ to the adapted coordinates $\{\alpha,\rho\}$ and
$t(\rho,\alpha)$ denotes the tensor insertion.  Terms with some of the cut
propagators missing are omitted in the cutting prescription. (One might extend
such terms with the necessary inverse propagators and see it vanish when the
$\rho$'s are set to zero.) 

For the \ibp{} vectors that do not double propagators we now show that
taking total derivatives commutes with the cut operation. That is, propagators
drop out of one class of terms in the total derivative, which leads to
vanishing terms when cuts are applied,
\begin{eqnarray}\label{onoffmap}
   && \int \left[\partial_i \left( \frac{u^i\, \mu\,  t }{\invprops }  \right)
   + \partial_b \left( \frac{u^b\, \mu\,  t }{\invprops }  \right) \right]
   \measure \nn \\
    &&= \int \left[ \frac{ \partial_i \big( f^i \, \mu\,  t \big)
    }{\dropinvprops }  + \partial_b \left( \frac{u^b\, \mu\,  t }{\invprops }
    \right)\right]  \measure\\
    &&\cut\quad   \int \partial_b \Big(u^b\, \mu\,  t   \Big) \cutmeasure\,.\nn
\end{eqnarray}
Here we use the relation $u^i=\rho^i f^i$ and we suppress the arguments for
better readability. (In the above equation the labels run as well over hat and
tilde values.)

Had we first cut and then used (the pull back of) the \ibp{} vector to obtain a
total derivative we obtained the same answer. The special form of the \ibp{}
vector fields makes this identification possible in the first place: 
since they are tangent vectors of the maximal-cut phase spaces they are well
defined intrinsically on the phase spaces and a pull back is well defined.

It is important to mention that the above reasoning did not involve the
$\alpha$-integration itself and is valid as well at the level of integrands (or
volume forms considering their Lie-derivatives \cite{Dirschmid}).

In general the cut of a vanishing integral does not need to give a surface term
on phase space, but might vanish, e.g. due to the choice of the physical
integration contour.

\subsubsection{On/Off-shell map} 

Relating on-shell and off-shell surface terms we can exploit intrinsic
properties of the on-shell phase spaces.
Every off-shell total derivative from special \ibp{} vectors gives one on
the maximal cut (\ref{onoffmap}). In formal terms, we obtain exact holomorphic
forms of maximal degree for each surface term.  
We have already seen (\sect{numbasis}) that the basis of tensor insertions
gives linear independent functions on the maximal-cut spaces. By multiplying
with the proper volume element we obtain a holomorphic forms of maximal degree.
Given that the coefficient functions are holomorphic and the forms are of
maximal degree they are closed; their outer derivative vanishes.  Thus surface
terms and tensors are given as intrinsic objects of the phase spaces.

Master integrands can be counted on-shell. The number of master integrands is
given by the number of independent tensor insertions modulo the number of
surface terms. On-shell this amounts to the number of closed modulo exact
holomorphic forms, that is a topological (global) property of the phase spaces.
Thus a topological property, in fact the number of half-maximal cycles, counts
master integrands. We will return to counting forms in
\sect{onshellconstruction}.

\section{Construction of off-shell surface terms}
\label{offshellconstruction}

We now turn to the main result: the construction of off-shell surface terms
which are a central ingredient for the numerical unitarity approach.  The
construction differs from the one at one-loop
level~\cite{OPP,NumUnitarity,RevUnitarity} which relied on tensor algebra and
symmetries of one-loop integrals. 
The present construction reproduces the one-loop results and applies also to
multi-loop topologies.  Put differently, we obtain a complete set of \ibp{}
relations which might be valuable beyond its use for the unitarity approach.

The presentation focuses here on planar two-loop topologies, but gives as
well non-planar surface terms. Higher-loop generalizations should work in
a similar way as we will indicate briefly (\sect{geointer}).
We consider the four-dimensional construction which yields surface terms that
involve the four-dimensional part of the loop momentum.  These terms are as
well surface terms in $D$ dimensions.  Furthermore at one-loop level the
four-dimensional numerators were recycled for the $D$-dimensional approach and we
believe the same construction works here.  Nevertheless, additional \ibp{}
vectors can be found beyond the four dimensional ones (see
\eqn{Ddimvecs}).  

The central objects are specialized \ibp{} vectors, which upon computing
divergences, are used to obtain the complete set of off-shell surface terms (see
\sect{sect:surfaceterms}).  The master integrals are obtained as convenient
tensor insertions in the complement of the surface terms.
The construction works topology by topology and reuses one-loop results in sub
topologies.
In order to introduce the key steps we start with a one-loop example in
\sect{ibponeloop}.
Next, we turn to the two-loop problem. A generating set of \ibp{}
vectors is obtained first in adapted coordinates in \sect{ibpvectorsadapted}
and, finally, in canonical notation in \sect{ibpvectorsstandard}.  The latter is
the main result of this article and has a very natural geometric interpretation
as we discuss in \sect{geointer}.

The set of planar generating \ibp{} vectors have the additional property that
they leave the integration measure invariant, i.e. their divergence vanishes.
This simplifies the computation of surface terms to taking directional
derivatives of irreducible numerator basis.  In this way an over complete
set of surface terms can be obtained which are equivalent to a complete set of
\ibp{} relations excluding doubled propagators. We verify the completeness
using on-shell techniques in \sect{onshellconstruction}.

\subsection{A one-loop example}
\label{ibponeloop}

The ingredients we need are an irreducible basis of tensor insertions,
\ibp{} vectors and the integration measure in order to compute total
derivatives.  It will turn out, that all \ibp{} vectors can be generated by a
set of primitive ones, which we have to consider in detail. Furthermore, it
turns out that the primitive \ibp{} vectors leave the volume element
and all propagators invariant. Under these circumstances the surface
terms are directly obtained by acting with \ibp{} vectors on the irreducible
tensor basis. Although these statements hold more generally, we will discuss
these steps for the  triangle integrals.

\subsubsection{Numerator tensors for triangles}

The four degrees of freedom of the loop momentum are parametrized in adapted
coordinates by three inverse propagators $\rho^{1,2,3}$ and two internal
coordinates $\alpha^{1,2}$ which are constrained by a single quadratic equation
(\ref{eqn:quadoneloop}),
\begin{eqnarray} \label{triquadeqn} c(\alpha,\rho)=\alpha^1\alpha^1+
\alpha^2\alpha^2+C(\rho)=0\,, \end{eqnarray}
with $C(\rho)$ given by scalar terms and inverse propagators.  The coordinates
are related to the loop momenta through the contractions $\alpha^a=(n^a,\ell)$
and the definitions of the inverse propagators. 

Irreducible numerators are given by polynomials in the $\alpha$-variables (see
\sect{numbasis}).  For standard power-counting we should consider at most cubic
powers of the loop momentum in triangle functions.  Thus the tensor insertions
$(\alpha^1)^{k_1}(\alpha^2)^{k_2}=(n^1,\ell)^{k_1}(n^2,\ell)^{k_2}$ with
$(k_1+k_2)\le 3$ suffice. Out of the ten monomials, only seven are linearly
independent modulo inverse propagators, as can be seen by using the quadratic
equation $c(\alpha,\rho)=0$, which relates inverse propagators and internal
coordinates. The three dependent monomials are $\sum_a(\alpha^a)^2\sim
-C(\rho)$ and $\alpha^{1,2}\sum_a(\alpha^a)^2\sim-\alpha^{1,2}C(\rho)$. 

It is convenient to make a linear coordinate change to the coordinates
$\alpha^\pm=(\alpha^1\pm i\alpha^2)$ (and $n^\pm=(n^1\pm i n^2)$), so that the
constraint equation is given by, 
\begin{eqnarray} \alpha^+\alpha^- +C(\rho)=0\,.  \end{eqnarray}
The reduction to a minimal numerator basis starting from monomials
$(\alpha^+)^{k_+}(\alpha^-)^{k_-}$ then simplifies; it amounts to dropping
mixed monomials in $\alpha^+$ and $\alpha^-$, since these are reducible
$\alpha^+\alpha^- \sim -C(\rho)$. The irreducible numerator basis is then given
by the seven terms $\{1,(\alpha^+)^m,(\alpha^-)^l\}$ with $l,k\le 3$ and thus the integrands,
\begin{eqnarray}
\frac{\tm_{\pm k}(\ell)}{\rho^1\rho^2\rho^3} 
= \frac{(\alpha^{\pm})^k}{\rho^1\rho^2\rho^3} =\,
\frac{(n^\pm,\ell)^k}{\rho^1\rho^2\rho^3}  \,.
\end{eqnarray}
We will
keep using the light-cone coordinates $\alpha^\pm$ below.

\subsubsection{Surface terms for triangles} 

The main ingredient for this construction are algebraic vector fields obeying
the condition (\ref{ibpcondition}). We start with the ansatz,
\begin{eqnarray} u=u^\pm\partial_{\alpha^\pm} + \sum_{i=1,2,3}f^i\rho^i
\partial_{\rho^i}\,.  \end{eqnarray}
As a simplification we set the function $f^i$ to zero and focus on such horizontal
vectors, which keep propagators fixed.  It turns out that this is no
restriction to obtaining a complete set of surface terms here.  Imposing the
consistency condition (\ref{algebraiccondition}) gives,
\begin{eqnarray}
 &&0=u\left(\alpha^+\alpha^-+C(\rho)\right)=u^+\alpha^-+u^- \alpha^+=0\,,\nn\\
 &&\Rightarrow \quad
 u=i(\alpha^+\partial_{\alpha^+}-\alpha^-\partial_{\alpha^-})\,,\nn
\end{eqnarray}
where we gave the simplest solution for the \ibp{} vector. The generic one is
obtained by multiplying $u$ with arbitrary polynomials in $\alpha$ and $\rho$
variables.

It is instructive to rewrite the obtained \ibp{} vector in canonical momentum
variables,
\begin{eqnarray}
u&=&i (\alpha^+\partial_{\alpha^+}-\alpha^-\partial_{\alpha^-})=
    (\alpha^{1}\partial_{\alpha^{2}}-\alpha^2\partial_{\alpha^1}) \nn\\
    && =\left[(n^1,\ell)  (n^2)^\mu - (n^2,\ell)
    (n^1)^\mu\right]\partial_\mu\,.
\end{eqnarray}
A number of remarks can be added here: 1) The \ibp{} vectors are
generators of rotations in the transverse space spanned by the $n^a$
vectors of the NV basis. 2) By construction the vector $u$ leaves the inverse
propagators invariant. In canonical coordinates this follows from the anti
symmetry of the vector and the property $(n^a,p_i)=0$,
\begin{eqnarray}
u(\ell^2)&=& 2\,\big[ (n^1,\ell)(n^2,\ell)-(n^2,\ell)(n^1,\ell)\big]=0\,,\nn\\
u((\ell-q_i)^2)&=& u\left( \ell^2 - 2 (\ell,q_i) \right) = 0 - 2\big[
(n^1,\ell)(n^2,q_i)-(n^2,\ell)(n^1,q_i)\big]=0\,.\nn
\end{eqnarray}
3) The divergence of the \ibp{} vector vanishes.  In canonical coordinates it
is straight forward to compute the divergence of a vector field, since the
volume element is constant\footnote{In general coordinate systems the
divergence of a vector field is obtained as its Lie-derivative
acting on the volume element \cite{Dirschmid}.} and one can verify,
\begin{eqnarray}
\partial_\mu u^\mu =0\,.
\end{eqnarray}
This property simplifies the computation of the divergences of generic \ibp{}
vectors which are obtained by multiplying the primitive ones by algebraic
tensors $t(\ell)$,
\begin{eqnarray}
\partial_\mu \left(t(\ell) u^\mu\right) = u^\mu \partial_\mu t(\ell) \,.
\end{eqnarray}
Thus the numerators of the triangle surface terms are directional derivatives
of generic tensors $t(\ell)$  with respect to the primitive \ibp{} vector.
Given that the $\rho$-components have been set to zero the propagator terms do
not have to be considered when computing the total derivative (or divergence). 

After these remarks it is straight forward to put together the \ibp{} relations of
surface terms. First of all, we write down a redundant set of \ibp{} vectors by
multiplying the generator $u$ with the irreducible numerator basis,
$(\alpha^\pm)^m u^\mu$. Then we calculate the divergences and obtain,
\begin{eqnarray}
\frac{\hm_{\pm k}(\ell)}{\rho^1\rho^2\rho^3} 
= \partial_\mu \left[ \frac{(\alpha^{\pm})^k u^\mu}{\rho^1\rho^2\rho^3} \right]
= \frac{u^\mu\partial_\mu (\alpha^{\pm})^k }{\rho^1\rho^2\rho^3} 
= \frac{\pm k\, (\alpha^{\pm})^k}{\rho^1\rho^2\rho^3} =\pm k\,
\frac{(n^\pm,\ell)^k}{\rho^1\rho^2\rho^3}  \,.
\end{eqnarray}
The only irreducible numerator that does not appear is the constant one
($k=0$), which is associated with the master integral. This is the well known
result for triangle integrals. Often surface terms are represented as symmetric
traceless tensors; the tensors $(n^\pm_\mu \cdot\cdot\cdot n^\pm_\nu)$ are
symmetric as well as traceless since the vectors $n^\pm$ have vanishing norm.

\subsection{Special two-loop \ibp{} vectors -- adapted coordinates}
\label{ibpvectorsadapted}

For a given topology we transform to the adapted coordinates and use the
observation (\ref{ibpconditionsol}) that the \ibp{} vectors have to take the
form (\ref{ibpcondition}) with their $\rho^i$-components being proportional to
the associated inverse propagators $\rho^i$. With this ansatz we impose the
conditions (\ref{algebraiccondition}) for algebraic vector fields. Given the
simplicity of the equations, we can solve them by inspection. 

\subsubsection{One-loop \ibp{} vectors}

We require first vectors that leave the quadratic equations $c=\tc=0$
of the individual loops invariant. Interesting \ibp{} vectors
are:
\begin{enumerate} \item[(a)] Generators of rotations,
\begin{eqnarray}\label{ibpgenerators}
    u_{[ab]}&=& \alpha_a\partial_b - \alpha_b \partial_a\,,
\end{eqnarray}
and analogously for the tilde-coordinates.  

\item[(b)] Vectors with non-trivial $\rho$-components,
\begin{eqnarray}
    u_{i}&=& \alpha^a \rho^i\partial_{\rho^i} - \rho^i(\partial_{\rho^i} c)
    \partial_a \,,
\end{eqnarray}
(no summation over the index $i$). We do not need this class of vectors
here. Nevertheless, it  would be interesting to understand the role of these
vectors better, as they may be used to relate tensor insertions of distinct
integral topologies. 

\item[(c)] $D$-dimensional vectors,
\begin{eqnarray}\label{Ddimvecs}
    u_{a}&=& \alpha^a \mu^b\partial_{\mu^b} - (\mu^b\mu^b)
    \partial_{\alpha^a}\,,
\end{eqnarray}
where the labels $b$ are summed over. Some explanations are required here.
Going beyond four dimensions the dimensionality of the loop momentum is
increased leading to additional transverse directions. The additional
transverse coordinates are conventionally called $\mu^b:=\alpha^{b+D_t}$.
Rotation-invariance in the $(D-4)$ dimensional directions is maintained and made
manifest for the above vectors. We give these vectors for completeness but will
not consider them further here.
\end{enumerate}
For our purposes only the vectors of type (a) will be important. We will refer
to such vectors, which do not contain components in the propagator directions
as horizontal vectors.

\subsubsection{Two-loop \ibp{} vectors}

Next, we consider the central rung and form linear combinations of the rotation
generators (\ref{ibpgenerators}) which leave the rung equation $\hc=0$
invariant:
\begin{enumerate}
\item[(a)] For more than two transverse variables in a given loop we find that the
rung-condition singles out a rotation axis~\footnote{Rotations are enumerated
by the inequivalent $(D-2)$-dimensional planes they leave invariant. We use the
term axis in an intuitive way to state that the set of planes is restricted to
the ones that contain the specified axis vector.} and we can write the linear
combination,
\begin{eqnarray}
    u_{[abc]}= e_{[a|}\,u_{|bc]} \,,\label{ibpvecaxis} 
\end{eqnarray}
where the notion $[abc]$  denotes the labels' anti-symmetric combinations. The
rotation axis is obtained by acting with the generator on the rung relation
(\ref{cond:rung}),
\begin{eqnarray}
    e_a= \partial_a \hc \,.
\end{eqnarray}
The anti-symmetric index structure makes sure that the rung condition is
annihilated by the vector (\ref{ibpvecaxis}).
\item[(b)] Diagonal rotations on the $\alpha$ space and $\talpha$ spaces give valid
\ibp{} vectors, whenever the internal spaces are lined up in at least two
directions and the rung relation is quadratic in the respective
$\alpha$-variables. This is the case in some semi-generic or simple topologies
(see \sect{eqntopology}).  (The coordinates in quadratic parts of the rung
relations must not appear in linear terms.) The vectors are then,
\begin{eqnarray}
    u^{\rm diag}_{[ab]}=  u_{[ab]} + \tu_{[ab]} \,.\label{ibpvecdiagonal} 
\end{eqnarray}
\item[(c)] Crossed rotations: these can appear, whenever both sides give rise to
at least two internal $\alpha$-coordinates each, 
\begin{eqnarray}
    u_{[ab][cd]}= \te_{cd} \, u_{[ab]} - e_{ab} \tu_{[cd]}
    \,.\label{ibpveccrossed} 
\end{eqnarray}
Here the anti-symmetric quantities $e_{ab}$ and $\te_{ab}$ are given by the
infinitesimal transformations of the rung equation,
\begin{eqnarray}
    \te_{ab}:=\tu_{[ab]} \hc\,,\quad \mbox{and}\quad  e_{ab}= u_{[ab]}\hc\,.
\end{eqnarray}
\item[(d)] Additional vectors can be constructed including inverse-propagator
derivatives as well as expressions including $D$-dimensional components. We will
not discuss these vectors here.
\end{enumerate}

Multiplying the above vectors with generic tensors gives further valid \ibp{}
vectors,
\begin{eqnarray}
\prod_{a,\ta} \alpha_a^{k_a} \,\talpha_{\ta}^{\tk_{\ta}} \times u,
\end{eqnarray}
which can be used to construct an over-complete list of surface terms. We will
discuss the construction of surface terms in more detail in
\sect{sect:surfaceterms}.

We refer to the vectors (a)-(c) as horizontal, given that they generate
motions that do not alter the propagator values.  These vectors turn into
tangent vectors of the unitarity-cut phase spaces, when on-shell conditions are
imposed. At the same time they may be viewed as an off-shell continuation of
the tangent bundles of the on-shell phase spaces to generic propagator values.
The continuation is not unique. Here the \ibp{} equation (\ref{fterms})
with a vanishing right hand side as well as the underlying Lie-algebra structure
allow one to control the off-shell continuation.

\subsubsection{Non-planar \ibp{} vectors}

The number of transverse variables limits the \ibp{} vectors that can be
introduced.  The relevant topologies are given by the $(N,\tN,\hN)=(2,2,2)$ and
$(N,\tN,\hN)=(2,2,3)$. All remaining non-planar topologies have no
unconstrained internal degrees of freedom and no surface terms with transverse
coordinates only may be obtained. 
The $(2,2,2)$-topology has six internal coordinates ($\alpha^{1,2,3}$ and
$\talpha^{1,2,3}$) which together with four constrains give a two dimensional
internal space. The $(2,2,3)$-topology start from five internal coordinates
($\alpha^{1,2,3}$ and $\talpha^{1,2}$) and remain with a one parameter
after all constraints are imposed. We base the construction on the
parametrizations inherited from the planar case (\ref{nonplanarrel}) where two
constraint from the central rung are imposed, $\hc=\hc'=0$. 

One type of vectors may be introduced for both topologies,
\begin{eqnarray}\label{nonplanvecI}
    u&=& e_{[\ta\tb]}e'_{[cd]}\,u_{[fg]} - e_{[cd]}e'_{[\ta\tb]}\,u_{[fg]} + e_{[cd]}e'_{[fg]}\,u_{[\ta\tb]} \nn\\
    && - e_{[\ta\tb]}e'_{[fg]}\,u_{[cd]} + e_{[fg]}e'_{[\ta\tb]}\,u_{[cd]} - e_{[fg]}e'_{[cd]}\,u_{[\ta\tb]} \,, \\
    && e_{[ab]}=u_{[ab]}(\hc)\,, \quad e'_{[ab]}=u_{[ab]}(\hc')\,.\nn
\end{eqnarray}
Here we use 'tilde' labels to distinguish data of the left and right loop.
For the $(N,\tN,\hN)=(2,2,2)$ topology a second such vector can be constructed
by exchanging all variables of left and right loop $a\leftrightarrow
\ta$.

Two additional primitive vectors may be constructed for the $(2,2,2)$-topology,
\begin{eqnarray} \label{nonplanvecII}
u&=& \te\, u_{abc} - e\, u_{\ta\tb\tc}\,,\\
u'&=& \te\, u'_{abc} - e\, u'_{\ta\tb\tc}\,,\nn
\end{eqnarray}
with the various symbols defined by,
\begin{eqnarray} 
u_{abc}&=& e_{[a|}\alpha_{|b|} \partial_{|c]}\,,\quad  u'_{abc}=
e'_{[a|}\alpha_{|b|} \partial_{|c]} \,,\\ 
e_{a}&=&\partial_a(\hc)\,,\quad
e'_{a}=\partial_a(\hc')\,,\quad 
e= e_{[a|}\alpha_{|b|} e'_{|c]}\,,\quad 
\te= e_{[\ta|}\alpha_{|\tb|} e'_{|\tc]}\,. \nn
\end{eqnarray}
and similarly for tilde-coordinates.

\subsection{Special two-loop \ibp{} vectors -- standard notation}
\label{ibpvectorsstandard}

The one-loop rotation generators are given in standard notation by,
\begin{eqnarray}
    u_{[kl]}= (\ell,n_{[k|})(n_{|l]},\partial) \,,
\end{eqnarray}
which allow to construct the generic two loop vectors by composition. This
vector matches the one in \eqn{ibpgenerators} up to a coordinate
change. This can be verified by comparing their action on the internal
coordinates and propagators.

\subsubsection{Two-loop \ibp{} vectors}

The above primitive \ibp{} vectors (\ref{ibpvecaxis})-(\ref{ibpveccrossed}) can
be given as well in canonical notation using the loop momenta $\ell,\tell$ and
$\hell$. The momenta are related by momentum conservation
$\hell=-(\ell-q_0+\tell-\tq_0+p_b)+\hq_0$, but to simplify expressions we use
the dependent momentum $\hell$. In the following we set the arbitrary shift
vectors to zero $q_0=\tq_0=\hq_0=0$ to simplify the expressions.

The two-loop vectors are given by the following three types:
\begin{enumerate}
\item[(a)] Rotation around an axis, 
\begin{eqnarray}\label{axisrotations}
    u_{[ijk]}=  (\hell,n_{[i|})(\ell,n_{|j|})(n_{|k]},\partial) \,,
\end{eqnarray}
or with tilde-expressions exchanged with non-tilde ones. Hidden in
$(\hell,n_{a})$ we have terms containing the contractions $(\tell,n_{a})$ which
can be rewritten using the completeness relation $\delta^\mu_\nu = \sum_{\ta} \tn^{\ta\mu}
\tn^{\ta}_\nu + \sum_{\ti} \tp_{\ti}^\mu \tv^{\ti}_\nu$ to give $\big[ (\tell,\tn_{\ta})
(\tn_{\ta}, n_a ) + (\tell,\tp_{\ti}) (\tv^{\ti}, n_a )\big] $. The later terms
contain the terms $(\tell,\tp_{\ti})$ and thus  off-shell information when
expressed through inverse propagators.
\item[(b)] Diagonal rotations,
\begin{eqnarray}\label{diagonalrotations}
    u_{[kl]}= (\ell,n_{[k|})(n_{|l]},\partial)+
    (\tell,\tn_{[k|})(\tn_{|l]},\tilde\partial)   \,.
\end{eqnarray}
\item[(c)] Crossed rotations,
\begin{eqnarray}\label{crossedrotations}
    u_{[ij][kl]}&=& (\tell,\tn_{[k|})(\tn_{|l]},\hell) \,
    (\ell,n_{[i|})(n_{|j]},\partial)-\nn\\ &&\quad\quad-
    (\ell,n_{[i|})(n_{|j]},\hell) \,
    (\tell,\tn_{[k|})(\tn_{|l]},\tilde\partial) \,.
\end{eqnarray}
\end{enumerate}
It follows from the symmetrization properties and the definitions of the
transverse space vectors, that all these vectors in fact annihilate inverse
propagators. This confirms that the listed vectors are valid specialized \ibp{}
vectors.

Multiplying the above vectors with generic
tensors gives further \ibp{} vectors,
\begin{eqnarray}\label{genericibpvec}
\prod_{i,j} (n_i,\ell)^{k_i} \,(\tn_{j},\tell)^{k_{j}} \times u,
\end{eqnarray}
which can be used to give an over-complete list of surface terms.  For the
planar topologies (\sect{topologies}) we have checked for that this list is
sufficient to generate all surface terms (see \sect{onshellconstruction}).

\subsubsection{Vectors and topologies}
\label{topologies}

The dimensionality and the alignment of the transverse spaces of a given
integral topology determines which of the above vectors are present. We collect
this information in \tab{vectortable}. We use the notation of \sect{intclass}
and label integral topologies by the pairs $(n,\tn)=(N-1,\tN-1)$. 

Type-$a$ vectors can be constructed, if the dimension of one-loop
transverse spaces at least three; $n<2$ or $\tn<2$.  Type-$b$ vectors can
be constructed, if the common transverse space is at least two-dimensional.
Vectors of type-$c$ can be constructed, if each of the transverse spaces is
at least two-dimensional; $n<3$ and $\tn<3$.

For example, the $(1,2)$-topology has a physical space spanned by
$\{p_1,\tp_1,\tp_2\}$. The left transverse space is $D-1=3$ dimensional and the
right transverse space $D-2=2$ dimensional.
For the 'generic' and 'semi-generic' topologies the external momenta
$\{p_1,\tp_1,\tp_2\}$ are linear independent and thus they span
three-dimensional space. The common transverse space is then one-dimensional.
We can construct type-$a$ vectors, since the left transverse space is
three-dimensional. The common transverse space is one-dimensional and no
type-$b$ vectors exist. Type-$c$ vectors exist, given that each loop has at
least two transverse directions. For the 'simple' $(1,2)$-topology the span
$\{p_1,\tp_1,\tp_2\}$ is only two-dimensional and the common transverse space
has two dimensions and type-$b$ vectors can be constructed.

\begin{table}[t]
\begin{center}
\begin{tabular}{ || c || c | c | c || }
      \hline
     & \multicolumn{3}{ c || }{ types of \ibp{} vectors}  \\\cline{2-4}
    (legs left,legs right)&   generic &   semi-generic  &  simple  \\\hline\hline    
    ( 0 , 0 )  & $a,b,c$    & --        &  --       \\\hline
    ( 0 , 1 )  & $a,b,c$    & $a,b,c$   &  --       \\\hline
    ( 0 , 2 )  & $a,b,c$    & $a,b,c$   &  $a,b,c$  \\\hline
    ( 0 , 3 )  & $a$        & $a$       &  $a$      \\\hline\hline
    ( 1 , 1 )  & $a,b,c $   & $a,b,c$   & $a,b,c$       \\\hline
    ( 1 , 2 )  & $a,c$      & $a,c$     & $a,b,c$       \\\hline
    ( 1 , 3 )  & $a$        & $a$       & $a$       \\\hline\hline
    ( 2 , 2 )  & $c$         & $c$       & $c$      \\\hline
    ( 2 , 3 )  & --         & --       & --      \\\hline
\end{tabular}
\caption{
\label{vectortable}
Here we display how the \ibp-vector types are associated with integral
topologies.
Three types of \ibp{} vectors are presented in \sect{ibpvectorsstandard}.
These are associated with rotations around an axis (type a), diagonal rotations
(type b) and crossed rotations (type c). The integral topologies determine
which and how many of the three types of vectors can be constructed. Here we
summarize this information for planar integral topologies. 
Only planar topologies of the master integrals with a central rung are
considered. The topologies are specified by (legs left,legs right); this refers
to the attached legs on the left/right loop (see \sect{intclass}).
The $(2,3)$ topologies have no unconstrained internal coordinates and, thus, no
horizontal \ibp{} vectors can be constructed.
} \end{center} \end{table} 

\subsubsection{Non-planar \ibp{} vectors}

As discussed above the relevant topologies are given by the
$(N,\tN,\hN)=(2,2,2)$ and $(N,\tN,\hN)=(2,2,3)$.

The first type of vectors may be introduced for both types of topologies. 
\begin{eqnarray}
    u&=& \te_{[ij]}e'_{[kl]}\,u_{[mn]} - e_{[kl]}\te'_{[ij]}\,u_{[mn]} + e_{[kl]}e'_{[mn]}\,\tu_{[ij]} \nn\\
    && - \te_{[ij]}e'_{[mn]}\,u_{[kl]} + e_{[mn]}\te'_{[ij]}\,u_{[kl]} - e_{[mn]}e'_{[kl]}\,\tu_{[ij]} \,,
\end{eqnarray}
with the auxiliary definitions,
\begin{eqnarray}
    && u_{[ij]}= (n_{[i|},\ell) (n_{|j]},\partial)\,,\quad \tu_{[ij]}= (\tn_{[i|},\tell) (\tn_{|j]},\tpartial)\,,\nn\\
    && e_{[ij]}=(n_{[i|},\ell) (n_{|j]},\hell) \,, \quad
        e'_{[ij]}=(n_{[i|},\ell) (n_{|j]},\hp_1) \,,\nn \\
  && \te_{[ij]}=(\tn_{[i|},\tell) (\tn_{|j]},\hell) \,, \quad
        \te'_{[ij]}=(\tn_{[i|},\tell) (\tn_{|j]},\hp_1) \,.\nn
\end{eqnarray}
We used the rung momentum $\hell$ which may be obtained from $\ell$ and $\tell$
by momentum conservation. The arbitrary shift vectors $q_0, \tq_0$ and $\hq_0$
are set to zero for simplicity.
For the $(N,\tN,\hN)=(2,2,2)$ topology a second such vector can be constructed
by exchanging all variables of left and right loop. 

Two additional primitive vectors can be constructed for the $(2,2,2)$-topology,
\begin{eqnarray} 
u&=& \te u_{ijk} - e \tu_{ijk}\,,\\
u'&=& \te u'_{ijk} - e \tu'_{ijk}\,,\nn
\end{eqnarray}
with the various symbols defined by,
\begin{eqnarray} 
u_{ijk}&=& (n_{[i|},\hell) (n_{|j|},\ell) (n_{|k]},\partial)\,,\quad  
u'_{ijk}= (n_{[i|},\hp_1) (n_{|j|},\ell) (n_{|k]},\partial)\,,\nn\\
\tu_{ijk}&=& (\tn_{[i|},\hell) (\tn_{|j|},\tell) (\tn_{|k]},\tpartial)\,,\quad  
\tu'_{ijk}= (\tn_{[i|},\hp_1) (\tn_{|j|},\tell) (\tn_{|k]},\tpartial)\,,\\
e&=&  (n_{[i|},\hell)(n_{|j|},\ell)(n_{|k]},\hp_1)\,,\quad 
\tilde e=  (\tn_{[i|},\hell)(\tn_{|j|},\tell)(\tn_{|k]},\hp_1)\,.\nn
\end{eqnarray}

\subsection{Surface from \ibp{} vectors for planar integrals}
\label{sect:surfaceterms}

The surface terms are obtained by computing the divergence with the above
\ibp{} vectors inserted into the loop integrands (\ref{genericibpvec}). 

A number of properties of the vectors lead to simplified expressions in the
planar topologies.  All vectors annihilate the inverse propagators.
Furthermore, the divergence of the above primitive \ibp{} vectors
(\ref{axisrotations}-\ref{crossedrotations}) vanishes: the total derivative of
rotation generators vanishes because of its antisymmetric index structure.
Similarly the diagonal rotation generators (\ref{diagonalrotations}) as well as
the ones around an axis (\ref{axisrotations}) have vanishing divergence.
Finally the crossed rotation generators (\ref{crossedrotations}) give, 
\begin{eqnarray} \label{divfree}
    \partial_\mu u_{[ij][kl]}^\mu + \tpartial_\nu \tu_{[ij][kl]}^\nu =
    (\tell,\tn_{[k|})(\tn_{|l]},n_{[j|}) \, (\ell,n_{|i]}) -
    (\ell,n_{[i|})(n_{|j]}, \tn_{[l|} ) \, (\tell,\tn_{|k]}) = 0 \,.
\end{eqnarray}

In general, inverse propagators
have to be included in computing surface terms from \ibp{} vectors
(\ref{surfaceterms}), however, with the \ibp{} vectors annihilating all inverse
propagators the derivation simplifies to taking the divergence,
\begin{eqnarray} \label{mainres}
     \hm_u &=& \partial_\mu\left(    \prod_{a,\ta} (n_a,\ell)^{k_a}
     \,(\tn_{\ta},\tell)^{k_{\ta}} \times u^\mu \right)
    +\tpartial_\nu\left(    \prod_{a,\ta} (n_a,\ell)^{k_a}
    \,(\tn_{\ta},\tell)^{k_{\ta}} \times \tu^\nu \right)\nn \\
    &&= \big( u^\mu\partial_\mu + \tu^\nu\tpartial_\nu\big) \, \left(
    \prod_{a,\ta} (n_a,\ell)^{k_a} \,(\tn_{\ta},\tell)^{k_{\ta}} \right)\,.
\end{eqnarray}
This is the main result of this article providing an over complete list of
surface terms; these are given by the action of the above \ibp{} vectors
(\ref{axisrotations}-\ref{crossedrotations}) on the basis tensors
(\ref{eqn:numbasis}). The master integrals have to be taken from the complement
and can be chosen as the lowest powers of the irreducible numerators.  In
\sect{onshellconstruction} we validate that this set of surface terms is in
fact complete. For the non-planar case the divergence-containing terms of the
primitive vectors are not expected to vanish, but can easily be included.

It is interesting to note here that numerators $\hm_u$ (\ref{mainres}) can be
inserted as well for topologies with the powers of the propagators altered,
\begin{eqnarray} \label{mainresp}
\int d^D\ell d^D\tell \frac{ \hm_u(\ell,\tell)}{\invpropsalter}=0\,,
\end{eqnarray}
for the horizontal vectors $u$ defined in \sect{ibpvectorsstandard} and
    \sect{nonplanparam}. 

\subsection{Geometric interpretation} \label{geointer}

\begin{figure}[t]
\includegraphics[clip,scale=0.80]{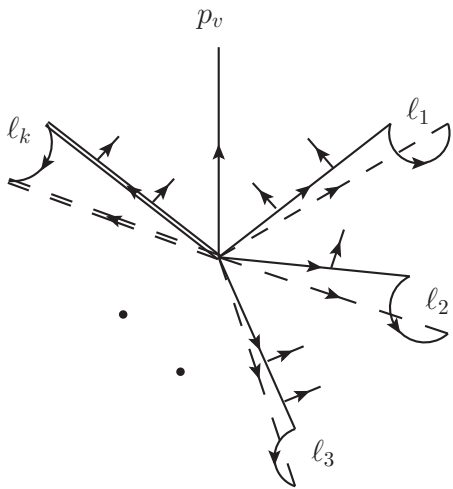} \caption{
    The junction of internal lines of a generic multi-loop diagram is
    displayed.  The loop momenta $\ell_i$ and the external momentum $p_v$ join
    at the vertex. The momenta are constrained by momentum conservation.  Each
    of the loop legs (here referred to as rungs) may have external momenta
    entering, which is indicated by small attached arrows. \ibp{} vectors
    generate rotations in the respective transverse spaces of the individual
    rungs.  Vertices impose interesting compatibility conditions between the
    rotations of the individual rungs.
    Via momentum conservation individual rotations of the rungs with loop
    momenta $\{\ell_1,...,\ell_{k-1}\}$ lead to a resulting transformation of
    $\ell_k$. The resulting transformation must be a rotation within this
    rung's transverse space to give a valid \ibp{} vector.}
\label{multirungFig} \end{figure}

The picture that emerges is that horizontal \ibp{} vectors are
particular infinitesimal rotations in the transverse spaces. In fact, we may
decompose a generic multi-loop diagram into rungs, which join in vertices as
displayed in \fig{multirungFig}.
%
%
For each rung we can use adapted coordinates, identify an associated transverse
space and, as in the one-loop problems, obtain simple quadratic equations
(\ref{eqn:quadoneloop}). Candidate \ibp{} vectors have to keep the individual quadratic
equation invariant and thus must be generators of rotations. (As above we
set propagator components of the vectors to zero.) At the vertices momentum
conservation has to be imposed yielding additional linear constraints. To
discuss this case further we assume a vertex that joins $k$ rungs with
additional external momentum $p_v$ entering,
$$\ell_{k}=-(\ell_1+ ... +\ell_{k-1}+ p_{v})\,.$$
Typical \ibp{} vectors have to be combinations of the individual rotation
generators, which rotate within the individual transverse spaces. 
Momentum conservation then imposes an interesting compatibility condition
between the rotations; it requires that the individual rotations of the
rung momenta $\ell_{i\neq k}$ lead to a resulting rotation in the transverse space of
the dependent momentum $\ell_k$. 

There are a number of interesting questions that should be addressed.  Maybe
the most fundamental being the question if it suffices to consider horizontal \ibp{}
vectors with vanishing $\rho$-components. Furthermore, it would be interesting
to take further advantage of the representation theory of the generators. For
example it should be possible to decompose polarization states with respect to
their rotation properties under the \ibp{} transformations in order to reduce
state sums in the loops.  We leave these and other questions to the future.

\subsection{Lie-algebra structure} \label{intstr}

We add a speculative discussion about the fundamental question, in how far the
physical amplitudes are determined by unitarity? We want to address this
question leaving the master integrals aside and consider their coefficients as
the physical quantities. 

In order to obtain the integral coefficients, we first try to reconstruct the
integrand. Given that perturbative field theories are algebraic and power counting
conditions hold, we have a finite number of terms which can be fixed on-shell on
generalized cuts.  Terms that are missed in one unitarity cut have to be
proportional to inverse propagators and can be determined if the set of all
generalized cuts is considered. Knowing the integrand, we still have to
identify the physical terms which correspond to the coefficients of the master
integrals. 

Given the loop integrand, the physical information can be extracted once we
have a split up into surface terms and master integrands; it is given by the
coefficients $d_i$  of the master integral.
The analogous statement holds for the numerators of unitarity cuts; once the
split up into surface terms (closed forms) and the remaining on-shell master
integrands is known, we can obtain the physical information of the unitarity
cut as the master coefficients $d_j^{\rm on-shell}$.

However, a prior, it is not clear that the same  physics is contained in
the two kinds of master coefficients, $d_i$ and $d_j^{\rm on-shell}$.  The core
structure for identifying master integrands are the above \ibp{} vectors in
their on-shell or off-shell versions. Each one of them determines a set of
surface terms and master integrals. Now, the fact that we find a correspondence
between the vectors, means that the on-shell and off-shell integrand
decompositions can be lined up and with it the physical coefficients. In this
way we believe that the off-shell continuation of the \ibp{} vectors is a
fundamental structure allowing the unitarity cuts to organise the amplitudes' physics,
possibly according to transcendentality.

We close this discussion with summarizing, that two structures play an
important role for the construction of the \ibp{} vectors: the \ibp{} relation
(\ref{fterms}) and the Lie-algebra structure. These do not only allow to construct the
vectors and surface terms but also provide a way to lift on-shell to off-shell
information.
The off-shell continuation works as follows: we first match the on-shell
vectors to rotation generators and then extend the generators off-shell
maintaining their Lie-algebra structure. The remaining ambiguity proportional
to inverse propagators is eliminated by considering additional unitarity cuts
and the horizontality condition.
    
\section{Counting master integrands on-shell} \label{onshellconstruction}

Here we use an on-shell approach to validate our main
results presented in the previous section.
To this end we compare two distinct ways to obtain master integrands for each of
the discussed integral topologies.  In the first approach we compare the linear
span of special \ibp{} relations (surface terms) with the span of irreducible tensor insertions. The
difference of the two gives the master integrands which we count,
\begin{eqnarray} \Nm=\Ni-\Ns\,.  \end{eqnarray}
We count modulo inverse propagators which is equivalent to counting the
functions when pulled back to the maximal-cut phase spaces.

In the second approach we consider the structure of on-shell phase spaces
directly to count master integrals.  The logic of the on-shell approach goes as
follows. 
We have shown in \sect{totdiffpullback} that all surface terms from \ibp{}
relations are turned into on-shell surface terms, that is exact forms on the
maximal cuts. Thus the set of all exact forms has to encompass the ones from
\ibp{} relations $\Ns \le \Ne$.
Furthermore, we have discussed in \sect{numbasis} that independent tensor insertions
remain independent functions on the maximal cuts.
Thus, when the tensor integrals are (maximally) cut, they yield linear
independent holomorphic forms of maximal degree on the maximal-cut phase
spaces $$\Ni=\Nc\,.$$

To exploit these observations we construct the complete set of exact forms of
maximal degree (a combinatorial problem) and then compare the linear spans of
exact and closed forms. The difference of the two is expected to give at least
a lower bound on the number of master integrals, $$\Nmp=\Nc-\Ne\,,$$
and is a topological property of the on-shell phase spaces.  We will confirm that we
obtain the same number of master integrals $\Nm=\Nmp$ in both approaches
(\sect{countmasters}). Given that the number of irreducible tensors matches in
both approaches, we conclude that we have found the maximal set of independent surface
terms $$\Ne=\Ns\,.$$ 
Thus, we verify that the set of primitive \ibp{} vectors is complete
and generates all surface terms.
Similarly, the set of surface terms is complete, in the sense that their complement in the
irreducible numerator tensors, i.e. the master integrands, are distinguishable
by unitarity cuts. Nevertheless, symmetric integration contours can still lead to
a small number of linear combinations of master integrands which integrate to
zero.

Throughout the construction we assume generic, non vanishing propagator and
external masses. This tries to accomplish two things: on the one hand this
mimics the $D$-dimensional questions and, on the other hand, the phase spaces of
the maximal cuts are then most regular allowing to use standard differential
calculus.

We have checked the approach for consistency. We have two distinct
software implementations producing identical results. We have anchored the
results with the examples we know of: we reproduce the (well known)
one-loop counts of master integrals and spurious terms. Furthermore, we
reproduce the count of nine master integrals in the double-box 
refs.~\cite{YZhangElliptic,YZhangElliptic2} and four master integrands
\cite{Tarasov97} in the sunset topology with generic masses. The results are
consistent with the integral count presented earlier in
ref.~\cite{SchabingerRadcor13} which, however, considers the four-point
topologies which often have vanishing external masses.

Some of the on-shell methodology has become available in a recent
publication~\cite{YZhangElliptic} for selected multi-loop topologies. These
involve typically maximal cuts yielding a one-dimensional phase spaces. We work
in a different direction considering generic multi-dimensional phase spaces,
exploit it to inspire the off-shell construction and compute the number of
master integrals for all planar integral topologies.

\subsection{A one-loop example}
\label{formexample}

We first explain the approach in a simple one-loop example namely the triangle
functions. The setup is analogous to the off-shell construction in
\sect{ibponeloop}. There the complete set of surface terms has been obtained as
well as the irreducible tensor insertions. We can read off the result,
$$\Nm=\Ni-\Ns=1 \,,$$
with $\Ni=7$ and $\Ns=6$.

We next turn to the on-shell approach.  Upon imposing the on-shell conditions
$\rho^i=0$, we are left with the on-shell phase space parameterized by
$\alpha^{\pm}$ which are constrained by the simplified quadratic equation
$c=\alpha^+\alpha^-+C(0)=0$.

\subsubsection{Algebraic function ring and differentials} 
The on-shell phase space is one-dimensional and we have to use
1-forms as integration measures. We first construct all 0-forms, i.e. algebraic
functions and then take outer derivatives to obtain the exact 1-forms.  The ring of
functions is generated by $\alpha^+$ and $\alpha^-$,
\begin{eqnarray}
(\alpha^+)^{k^+} (\alpha^-)^{k^-}\,,\quad k^++k^-\le 3\,,
\end{eqnarray}
with the rank of the monomials restricted by power counting.

The independent functions are obtained by using the relation
$\alpha^+\alpha^-=-C$ with the shorthand notation $C:=C(0)$. Thus, as in the
off-shell case the independent functions are given by $1$, $(\alpha^+)^m$ or
$(\alpha^-)^n$; whenever a mixed term of $\alpha^+$ and $\alpha^-$ appears it
can be turned into the constant $C$.
The set of all differentials are generated by the freely generated ring of
formal expressions $d\alpha^+$ and $d\alpha^-$
\begin{eqnarray} \label{example:ring}
(\alpha^+)^{k^+} (\alpha^-)^{k^-} d\alpha^{\pm}\,,\quad k^++k^-\le 3\,.
\end{eqnarray}
The linear independent differentials are obtained by imposing the on-shell
conditions $dc=0$ and $c\,d\alpha^{\pm}=0$ for the differentials. We can first
reduce to fourteen terms $(\alpha^\pm)^kd\alpha^\pm$ reducing the coefficient
functions.  Then we use the relation, $\alpha^+d\alpha^-+\alpha^-d\alpha^+=0$
to relate the forms $(\alpha^+)^{(k>1)}d\alpha^-=C\,(\alpha^+)^{k-2}d\alpha^+$
(and $(\alpha^-)^{(k>1)}d\alpha^+=C\,(\alpha^-)^{k-2}d\alpha^-$) to obtain the
seven independent forms ($\Nc=7$),
\begin{eqnarray}\label{exampleclosedforms}
(\alpha^+)^{k^+} d\alpha^{+}\,,\quad (\alpha^-)^{k^-} d\alpha^{-}\,,\quad
\alpha^+\, d\alpha^{-}=-\alpha^-\, d\alpha^{+}\,, \quad k^{\pm} \le 3\,.
\end{eqnarray}

The set of all exact differentials is given by taking outer derivatives on the
function ring (\ref{example:ring}) with $(k^++k^-)\le4$. Losing the constant function we obtain six
($\Ne=6$) exact forms $d((\alpha^\pm)^{k})=k\,(\alpha^\pm)^{k-1}d\alpha^\pm$.
These forms are linear independent and no further reduction is needed.
Compared to the set of all 1-forms the exact ones do not include $\alpha^-
d\alpha^+$ making it a representative of a master integral as the only
non-exact form. Thus we obtain,
$$\Nmp=\Nc-\Ne=1\,,$$
with $\Nc=7$ and $\Ne=6$. In more general cases it is often convenient to count
the combined sets of closed and exact forms as well as the exact forms, with a
relaxed power-counting restriction on the exact forms to avoids boundary
effects.
We observe $\Nm=\Nmp=1$ which is the well known result.
From the topology of the maximal-cut phase space this is expected,
since we have one non-trivial cycle to which the form $\alpha^+d\alpha^-$ is
dual to. 

We close with a remark: we can make contact with the standard on-shell notation: we
may solve the quadratic equation $\alpha^+\alpha^-=-C$ and use the explicit
parametrization $\alpha^+=t$ and $\alpha^-=-C/t$. The set of tensor insertions
is given by $t^k$ and $(C/t)^k$. It is straight forward to write down all
closed forms by acting with the outer derivative on the functions $t^k$
with $k$ being a positive or negative integer. From the generated forms
$(t^k\,dt/t)$ only $dt/t$ cannot be obtained from acting on the functions $t^k$
since $dt/t=d\,ln(t)$ with $ln(t)$ being non-algebraic.  Again we obtain a
single non-trivial closed form. We conclude that there is one master integrand
with the on-shell representation $dt/t$ and six surface terms $t^k \,dt/t$
with $k\neq0$ and $|k|\le3$.

\subsection{Two-loop computations}
Computing the off-shell data is straight
forward following the instructions of \sect{sect:surfaceterms}.
For the on-shell construction a number of ingredients are needed. We need to
give a classification of the relevant integral topologies and their on-shell
equations. Furthermore, we need an algorithm to construct exact and closed
forms. We will turn to these points in the following.

\subsubsection{Setup}
The maximal cuts are obtained using the adapted coordinates (\sect{LoopMomParam}) by
setting the inverse propagators to zero. Consequently, the maximal-cut phase spaces are
parameterized by the internal coordinates $\alpha^a$ and $\talpha^{\ta}$ with
quadratic equations (\ref{symequations},\ref{rungrelation}) of the following form imposed,
\begin{eqnarray} 
    &&c=\ax^a \ax^a + C = 0 \,, \nn\\
    &&\tc=\tax^{\ta}\tax^{\ta} + \tC = 0 \,, \nn\\
    && \hc=\C_{a\ta} \ax^a \tax^{\ta} + 
       \C_{\ta} \tax^{\ta} +
       \ax^a \C_{a}+ \C = 0 \,,\label{rungquadeqn}
\end{eqnarray}
where $C,\tC,\C,\C_a,\C_{\ta}$ and the matrices $\C_{a\ta}$ are
determined from the explicit kinematic configuration of external momenta and
the integral topology. In the non-planar case we had a further equation $\hc'=0$
but we focus on the planar cases now.

The function space is generated (non-minimally) by tensor insertions evaluated
on the maximal cut, which is given here by the monomials,
\begin{eqnarray}
   \prod_{a,\ta} \alpha^{k_a}_a\talpha^{\tk_{\ta}}_{\ta}  \,.
\end{eqnarray}
The set of independent (irreducible) functions is obtained by imposing the on-shell
conditions (\ref{rungquadeqn}). Algebraically one considers the above monomials
modulo multiples of the on-shell conditions; that is modulo the ideal generated
by the on-shell conditions.  We assume a fixed set of external momenta and
masses.  The coefficients can be viewed as constants and we can use
sufficiently generic integer-valued expressions for simplicity. 

Power counting restricts the monomials. Assuming that a vertex contributes at
most a single power of loop momentum we obtain the restriction $\sum_a k_a \le
(N+1)$, $\sum_{\ta} \tk_{\ta} \le (\tN+1)$ and $\sum_a k_a + \sum_{\ta}
\tk_{\ta} \le N+\tN$. For simplicity we will impose the power-counting
constraint in the following, although the approach is not limited to
power-counting renormalizable theories.


\subsubsection{Equations and topology}
\label{eqntopology}

The quadratic equations (\ref{rungquadeqn}) simplify for certain topologies.
This in turn changes the symmetry content, which is characterized by the \ibp{}
vectors, and consequently the number of master integrals. The final form of the
quadratic equations will be given below for the individual topologies. 

The first two quadratic equations are invariant under orthogonal transformation
of the internal space variables $\ax^a$ and $\tax^{\ta}$. These rotations may
be used to transform the rung-equation to a canonical form.  Here we use the NV
vectors to fix the form of the rung equation and insert integer coefficients for
our computations.

We distinguish integral topologies as well as integrals with and without
momenta ($p_b$ and $p_t$ in our conventions) attached to the top and bottom
rung vertices. These external momenta in the rung equations influence the
alignment of the transverse and physical spaces of the respective loops.  Three
classes of integrals are distinguished, where we restrict ourselves to the
diagrams with $n=(N-1)$ legs and $\tn=(\tN-1)$ legs attached to left and right
loop, respectively. Without loss of generality we assume $(\tn \ge n)$
configurations. The dimensions of the left and right transverse spaces are
$(D-n)$ and $(D-\tn)$ respectively. Apart from the distinct dimensionality we
assume that the transverse spaces differ by a generic rotation from one another.
\begin{enumerate}
\item Generic case: these correspond to ($n$, $\tn$)-leg diagrams with
non-vanishing $p_b$ and $p_t$. The dimension of the common physical space is
Min$(n+\tn,D)$ and the common transverse space has dimension Max$(D-n-\tn,0)$.
The quadratic equation for the central rung is given by,
\begin{eqnarray}
     &&\sum_{i=1}^{D-n-\tn} (\ax-\C)^{i} (\tax-\C)^{i} + 
     \sum_{b,\tb = D-n-\tn+1}^{D-\tn} \C_{b\tb} \ax^b \tax^{\tb}  +\nn\\ &&\quad 
    +\sum_{\tb = D-n-\tn+1}^{D-\tn} \C_{\tb} \tax^{\tb} + 
    \sum_{b = D-n-\tn+1}^{D-n } \ax^b
    \C_{b}+ \C = 0 \,.\nn
\end{eqnarray}
In the $(D-n-\tn)$-dimensional common transverse space we can line up the basis
vectors $n_i$ and $\tn_i$ to obtain a unit diagonal bilinear term; this
justifies the form of the first term. 
We have $\tn$ remaining transverse directions from the left loop and $n\le \tn$
from the right loop. The left transverse basis can then be rotated, so that
$(\tn-n)$ transverse vectors $n_{b>(D-\tn)}$ of the left loop are orthogonal to
the $\tn$ basis vectors $\tn_{b}$ of the right loop. The $n$ directions
$\{n_b,\tn_b\}$ with $(D-n-\tn+1)\le b \le (D-\tn)$ lie at generic angles with
respect to each other. The latter gives rise to the generic $n\times n$ matrix
$\C_{b\tb}\sim (n_b,\tn_{\tb})$.
The linear terms represent the orientation of the external momenta with respect
to the transverse bases. The vectors $\C^i$ arise from the inner products
$(p_b,n_i)=(p_b,\tn_i)$.  The constant vectors $\C_{b}$ and $\C_{\tb}$ arise
from contractions of the transverse space vectors of the left/right-loop with
the physical directions.
\item Semi-generic case: by this topology we mean that only one rung vertex has
external momentum attached. Without loss of generality we assume $p_b=0$. 
\begin{eqnarray} 
    \hspace{-1 cm} \sum_{i=1}^{D-n-\tn} \ax^{i} \tax^{i} + \sum_{b,\tb =
    D-n-\tn+1}^{D-\tn} \C_{b\tb} \ax^b \tax^{\tb} 
    +\sum_{\tb = D-n-\tn+1}^{D-\tn} \C_{\tb} \tax^{\tb} + \sum_{b = D-n-\tn+1}^{D-n } \ax^b
    \C_{b}+ \C = 0 \,.\nn
\end{eqnarray}
The only change compared to the generic case is that the constant
vectors $\C^i$ vanish since $p_b=0$.
\item Simplest case: by this topology we mean that $p_b=p_t=0$. Relative to the
generic case this implies the following simplifications. The vectors $\C^i$
vanish and, furthermore, the external momenta attached to the left and right
loop are linearly dependent. Thus the dimension of the common transverse space
increases by one,
\begin{eqnarray} 
    \hspace{-1.5 cm}\sum_{i=1}^{D-n-\tn+1} \ax^{i} \tax^{i} + \sum_{b,\tb =
    D-n-\tn+2}^{D-\tn} \C_{b\tb} \ax^b \tax^{\tb}  +\sum_{\tb =
    D-n-\tn+2}^{D-\tn} \C_{\tb} \tax^{\tb} + \sum_{b = D-n-\tn+2}^{D-n } \ax^b
    \C_{b}+ \C = 0 \,.\nn
\end{eqnarray}
\end{enumerate} 
These equations represent the essence of the two-loop integrands and give a
moduli space of the on-shell conditions. Here the equations are required as the
input for the computation of the exact forms as well as the irreducible
numerators.

\subsubsection{The algorithm}

We will now construct holomorphic differential forms of maximal degree on the
unitarity-cut phase spaces and relate them to surface terms obtained as \ibp{}
relations. We will compare and count both results as a consistency check and a
validation of the completeness of the surface terms computed earlier.

The considerations go as follows. Tensor insertions appear as volume elements
(or forms) on the internal space following the split up
(\ref{motivationcohomology}).  Furthermore, although the coordinates
$(\ax^a,\tax^{\ta})$ are complex valued the volume element arises from $t(\ell,\tell)\,d^D\ell
d^D\tell$, which is holomorphic so that it suffices to consider holomorphic
forms and functions.  (In practice this means that we do not need to use
$\bar\alpha^a, d\bar\alpha^a$ etc.) Since the holomorphic forms have maximal
degree they are in fact closed.

In addition to the closed forms we need to consider exact forms given that all
\ibp{} vectors generate exact forms on-shell (\sect{totdiffpullback}).  
Both types of forms are defined on the on-shell spaces which are algebraic
varieties. It will be sufficient to consider algebraic forms on these spaces
(see also \sect{dropmeasure}).

Eventually we count master integrals. These are given by closed forms that are
non-exact, i.e. that are not surface terms.

We will first generate over complete sets of closed and exact forms in a
combinatorial way. These sets of forms will be denoted by $\hOc$ and $\hOe$,
respectively.  In a second step we will reduce $\hOc$ and $\hOe$ by
vanishing forms $\Oz$ to obtain the linear independent sets $\Oc=\hOc/\Oz$ and
$\Oe=\hOe/\Oz$.  The vanishing forms $\Oz$ have nontrivial algebraic
expressions but vanish when the on-shell conditions are taken into account,
e.g. the one-form $dc$ is equivalent to zero on-shell $dc\sim 0$
(\ref{onshellrelationforms}). The form degree matches the dimensionality of the
phase space which we denote by $m$.  For convenience we label the on-shell
conditions by an index $i$; $\{c_i\}=\{c,\tc,\hc,...\}$ and enumerate the
transverse coordinates by a single label $i$ with
$\{\alpha^i\}=\{\alpha^a,\talpha^{\ta}\}$.

Algebraic $n$-forms can be obtained in a combinatorial way; first of all one
has to take outer products of the independent algebraic 1-forms to obtain
a basis of $n$-forms. Next one multiplies these forms by monomials in the
transverse coordinates to obtain forms up to a fixed polynomial rank. Such 
$n$-forms are given by, 
\begin{eqnarray} \label{formdef}
\omega_n= \left(\prod_{i} (\ax^{i})^{k_i} \right)d\ax^{j_1}\wedge \dots \wedge
d\ax^{j_{n}} \,. 
\end{eqnarray}
The symbols $k_i$ denote non-negative integers. Typically we are interested in
the linear span of such forms over the complex numbers.

The vanishing form $\Oz$ are given as the linear span of the following
$m$-forms:
\begin{enumerate}
\item Algebraic $m$-forms $\omega_m$ multiplied by an on-shell relation $c_i$, 
\begin{eqnarray}
    \omega_{zero}= c_i \,\, \omega_m \,.
\end{eqnarray}
\item Forms obtained from outer products with a differential of the on-shell
relation,
\begin{eqnarray}
    \omega_{zero}=dc_i\wedge \omega_{m-1} \,.
\end{eqnarray}
\end{enumerate}

The exact forms $\omega_{exact}\in\hOe$ are generated by taking outer derivatives on holomorphic forms
of one degree less than the volume form,
\begin{eqnarray}
    \omega_{exact}=d\omega_{m-1}\,.
\end{eqnarray}
The closed forms $\hOc$ are given by the holomorphic algebraic $m$-forms (\ref{formdef}).

With closed, exact and vanishing forms available, the independent forms are
obtained as equivalence classes of the relations,
\begin{eqnarray}
    \omega_{m}\sim \omega_{m}+\omega_{zero}\,.
\end{eqnarray}
We thus obtain $\Oc=\hOc/\Oz$ and $\Oe=\hOe/\Oz$.

Finally, we can count master integrals $\Om$ as closed forms that are
non-exact. We compare the linear spans of $\Oc$ and $\Oe$; we quotient $\Oc$
by $\Oe$, $$\Nmp=\dim(\Om)\,,\quad  \Om= \Oc/\Oe\,.$$
In principle the polynomial order of the forms is restricted by power counting.
However, it is often simpler to relax the power counting conditions. At least in
the present cases the number of master integrals is not reduced by power
counting conditions.
Furthermore, it is convenient to compute the dimensionality of $\Om$ indirectly
by 
\begin{eqnarray}
\Nmp=\dim(\Oc\cup\Oe)-\dim(\Oe)\,.
\end{eqnarray}
This helps to avoid boundary effects due to limiting the polynomial rank of the
forms; for a fixed polynomial degree the set of exact forms does not need to be
subset of the closed ones.

We implemented the above computations by translating to a related
polynomial problems (see \sect{formtechnique}). In fact we can treat the basis
forms as auxiliary coordinates. Exact as well as closed forms then
appear as polynomials.  The vanishing forms can be interpreted as algebraic
relations generating an ideal. The constructions of exact and closed forms then
transforms to reducing the polynomials associated with exact and closed forms in
the extended affine coordinate ring by the ideal of the vanishing forms.

\subsubsection{Discussion of the on-shell measure}
\label{dropmeasure}

The counting of master integrands in the previous subsections was based on
algebraic differential forms on the unitarity-cut phase space. However, while
the off-shell volume element is algebraic, the on-shell one arises from
integrating out delta-function constraints. Thus the cutting procedure leads to
holomorphic forms that differ from algebraic ones. For the smooth spaces we are
dealing with, the volume element carries a non-singular factor. (This factor can
be obtained from the coordinate changes but we will not need it here.) 

Nevertheless, we argue that the algebraic count provides sufficient
information.  The number of master integrals is given by a topological quantity
and we assume that the algebraic computation is sufficient to obtain its value.


\subsection{Master integrand counting} \label{countmasters}

We have implemented the construction of closed and exact algebraic differential
forms in \Mathematica{}~\cite{Mathematica}. As input we used the on-shell
conditions given in \sect{eqntopology} and we computed the expected number of
master integrals $\Nmp$. The computations were performed for generic numerical
coefficient matrices in the quadratic equations.

In a second computation we constructed the primitive \ibp{} vectors using
adapted coordinates. We then computed surface terms using these vectors.
(Technically we computed Lie-derivatives of the closed algebraic forms with
respect to the \ibp{} vectors.) We confirm that the number of master integrals
$\Nm$ matches in both approaches,
\begin{eqnarray} \Nmp=\Nm\,, \end{eqnarray}
for all considered planar topologies. As a reference we give the counts of the
    master integrands below in \tab{mastertable}.

For completeness we give as well the total number of irreducible numerator
tensors, which were obtained here by reducing the function ring of the internal
coordinates by the quadratic relations \sect{eqntopology}. These numbers match
the results in the literature~\cite{MLoopAlgGeo,MYZhangElliptic,LoopCohom,NumeratorsFH}.  This
check is based on the same input equations as used for computing surface terms
and counting master integrands. Thus comparing irreducible numerators validates
our input equations.

We included as well results for diagrams with bubbles on internal lines. These
topologies come with doubled propagators to start with. The corresponding
results are marked with an asterisk and referred to as 'simple' $(0,n)$
topologies. In the 'simple' topologies no legs are attached to the vertices of
the central rung. 

\begin{table}[t]
\begin{center}
\begin{tabular}{ || c || c | c | c || c || }
      \hline
     & \multicolumn{3}{ c || }{\# master integrands} & \\\cline{2-4}
    (legs left,legs right)& generic &   semi-generic  &  simple  & irred. tensors \\\hline\hline    
    ( 0 , 0 )  & $4$    & --    &  --   &   $42$    \\\hline
    ( 0 , 1 )  & $4$    & $1$   &  --   &   $80$    \\\hline
    ( 0 , 2 )  & $6$    & $1$   & $1^*$ &   $65$    \\\hline
    ( 0 , 3 )  & $2$    & $2$   & $2^*$ &   $18$    \\\hline\hline
    ( 1 , 1 )  & $9$    & $5$   & $1$   &   $111$   \\\hline
    ( 1 , 2 )  & $8$    & $8$   & $1$   &   $69$    \\\hline
    ( 1 , 3 )  & $2$    & $2$   & $2$   &   $14$    \\\hline\hline
    ( 2 , 2 )  & $9$    & $9$   & $9$   &   $32$    \\\hline
    ( 2 , 3 )  & $4$    & $4$   & $4$   &   $4$     \\\hline
\end{tabular}
\caption{
\label{mastertable}
The number of master integrands are given. No symmetry properties of the
integrals are taken into account.  Only planar topologies of the master
integrals with a central rung are considered. The topologies are specified by
(legs left, legs right); this refers to the attached legs on the left/right
loop and is denoted as well by $(N-1,\tN-1)$ in the text.
The number of master integrands are obtained in our on-shell approach and
validated using off-shell \ibp{} vectors.  The counts refer to integrals with
standard power counting of QCD like theories but are expected to hold as well
beyond this (e.g. gravity theories).  
For completeness, the total number of irreducible tensor insertions for the
given topologies is displayed. The numbers hold for QCD-like power counting.
The asterisk '$^*$' marks topologies with bubble diagrams on internal lines
which have doubled propagators to start with.
} \end{center} \end{table} 

The count of the master integrals is useful by itself but,
furthermore, confirms the completeness of the \ibp{} vectors given. Actually,
even more directly the count confirms that we have a complete set of surface
terms. Here complete refers to the property that the master integrands are
distinguishable by unitarity cuts.

The number of integrands is consistent with partially available
results~\cite{YZhangElliptic,MLoopCohom,SchabingerRadcor13}.  Comparing the number of master
integrals is less simple, given that the results for integrals and integrands
can differ (due to symmetry properties). Nevertheless, we find the results
consistent with available results. 

\section{Summary and future directions}

We have discussed new methods and results which are required for multi-loop matrix
element generators based on a numerical unitarity approach.
The methods form a synthesis of established techniques to obtain integral
relations (or surface terms) and the unitarity approach. Interesting
connections have been exposed between the classification of loop integrands,
special \ibp{} relations \cite{KosowerIBP} and topological properties of
unitarity-cut phase spaces.

We presented a number of results: First of all, we constructed a complete set
of off-shell integral relations (surface terms) (\ref{mainres}) for the planar
two-loop topologies. 
These relations are obtained from a new type of horizontal \ibp{} vectors. We
show that their associated \ibp{} relations hold as well if propagator
powers are simply changed (\sect{sect:surfaceterms}).
In addition, as an intermediate result we computed the number of master
integrands which is related to a topological property of the unitarity-cut
phase spaces.

Further results include the geometric interpretation of \ibp{} vectors and
convenient coordinates to identify irreducible tensor insertions
(\sect{numbasis}). 
Moreover, we exposed an important link between on-shell and off-shell
integrals of loop amplitudes (\sect{totdiffpullback}), and identify a
useful Lie-algebra structure in the construction
(sections \ref{ibpvectorsadapted}, \ref{ibpvectorsstandard} and \ref{intstr}).

We performed a number of checks. The results for the number of master
integrands are consistent with the partially available
results~\cite{YZhangElliptic,MLoopCohom,SchabingerRadcor13}.
Using the identical input, we count irreducible numerator tensors. The counts match
results from earlier constructions of loop integrands
\cite{MLoopParamMO,MLoopParamBFZ,MLoopAlgGeo,NumeratorsFH}.  
We obtained surface terms in two distinct ways; an explicit construction using
a generating set of \ibp{} vectors as well as a combinatorial, on-shell
approach. The fact that the results agree demonstrates that a complete set of
surface terms was obtained. This implies as well that the classification of the
generating \ibp{} vectors is complete for the generic planar integrals. 

An number of formal questions have to be addressed for computing multi-loop
amplitudes based on the presented methods.  First of all, it will be important
to verify that the provided \ibp{} vectors suffice for integral topologies with
vanishing external masses. This is plausible from factorisation properties but
needs to be checked in detail.
Second, it will be interesting to understand the role of horizontal \ibp{}
vectors in higher-loop computations. These vectors are an off-shell
continuation of the tangent vectors of the unitarity-cut phase spaces and
should continue to be the core structure for the construction of multi-loop integrands.
%
%
%
Third, we exposed a Lie-algebra structure, given that the \ibp{} vectors are
generators of rotations. It will be useful to exploit this to simplify
computations further (e.g. reduce state sums). 
Furthermore, with generic expressions given, one may study the integrands'
dependence on external momenta (e.g. obtain differential equations for master
integrals). 
Finally, interesting extensions include as well the application of the surface
terms to the computation of real pieces and subtraction terms.

The best combination of analytical and numerical methods for multi-loop
computations is not obvious in the moment, however, we are motivated by the
one-loop successes of the unitarity method to aim for analogous strategy for
multi-loop computations.  
%
%
The computation of NNLO QCD cross sections is a very challenging task given that so many
pieces have to be controlled and combined. We hope that the methods discussed
in this article can contribute to new precision predictions for the LHC
experiments on the long run.

\section*{Acknowledgments}
I would like to thank S.~Abreu, Z.~Bern, A.~Brandhuber, F.~Febres Cordero, 
M.~Jaquier, K.~Larsen, B.~Page, F.~Schlenk and M.~Zeng for many helpful discussions and comments. 
Furthermore, I would like to thank L.~Dixon, D.~Forde, D.~Kosower, D.~\Maitre{}
and K.~Ozeren for many inspiring discussions on related topics.
This work is supported by a Marie Sk{\l}odowska-Curie Action Career-Integration
Grant PCIG12-GA-2012-334228 of the European Union and by the Juniorprofessor
Program of Ministry of Science, Research and the Arts of the state of
Baden-W\"urttemberg, Germany.  

\appendix

\section{Technical construction of algebraic forms}
\label{formtechnique}

We can relate the problem of constructing differential forms on smooth
algebraic surfaces to reducing an affine coordinate ring by an ideal.
We have already encountered the reduction of a  polynomial ring by an ideal when
constructing irreducible numerators (\sect{numbasis}): to this end we write
down all polynomials of the internal $\alpha$-variables and use the on-shell
relations $c=\tc=\hc=0$ to identify equivalent polynomials.
The construction of forms is related to this. We start with the standard form
calculus in the affine space and consider differential forms with polynomial
coefficient functions. When we pull back the forms to the maximal-cut phase
spaces, initially independent forms become linearly dependent. In order to reduce
the linearly dependent forms we use a technical trick; we collect the basis
forms as well as the affine coordinates into an extended coordinate
ring (see. \cite{Hartshorne} Chapter II.8 on Differentials for inspiration),
\begin{eqnarray}
    \mbox{coordinate ring:}\quad \{\ax^1,\ax^2,  ... , y^1=d\ax^1,y^2=d\ax^2,
    ... \}  \,.
\end{eqnarray}
When pulled back to the maximal-cut phase space, combinations of the forms
vanish, which is expressed by an ideal in the auxiliary space of $\alpha$ and
$y$ coordinates,
\begin{eqnarray}
    \mbox{generators of ideal:}\quad \{c,\tc,\hc,  ... , dc ,d\tc,d\hc, ... , y^i y^j, ... \}
    \,.
\end{eqnarray}
Here the differentials of the on-shell relations $dc=(\partial_{a} c)\, d\ax^a
=(\partial_{a} c)\, y^a$ give polynomial equations mixing $\alpha$'s and $y$'s.
Furthermore, the relations $y^i y^j=0$ constrain us to form-degree one.  The
set of linearly independent forms is then obtained by reducing polynomials in
$\alpha$'s and $y$'s which are linear in $y^b$ by the ideal.

For forms of generic degree ($n$) one proceeds analogously. Again one includes
all basis forms as coordinates,
\begin{eqnarray}
    \mbox{coordinate ring:}&& \{\ax^1,\ax^2,  ... ,
    y^{[i_1,...,i_n]}=d\ax^{[i_1,i_2,...,i_n]}, ... \}  \nn\\ &&
    \mbox{with}\quad d\ax^{[i_1,i_2,...,i_n]}:= d\ax^{i_1}\wedge ... \wedge
    d\ax^{i_n} \,, \nn
\end{eqnarray}
and defines relations (an ideal) by the original on-shell relations, their
derivatives as well as squares of the differentials,
\begin{eqnarray}
    \mbox{generators of ideal:}&&  \{c,\tc,  ... , dc\wedge d\ax^{[i_1,i_2,...i_{n-1}]}
    ,..., d\tc\wedge d\ax^{[i_1,i_2,...i_{n-1}]}, ...  ,\nn\\&&
    y^{[i_1,i_2,...,i_n]}\,y^{[j_1,i_2,...,j_n]}, ... \}  \,, \nn
\end{eqnarray}
where the expressions $dc\wedge d\ax^{[i_1,i_2,...i_{n-1}]}$ have to be written
as linear expressions in the coordinates $y^{[i_1,...i_n]}$,
\begin{eqnarray}
     dc_j\wedge d\ax^{[i_1,i_2,...,i_{n-1}]}&=&\partial_{b}c_j\, d\alpha^b\wedge d\ax^{[i_1,i_2,...,i_{n-1}]}
     = (\partial_{b}c_j)\, y^{[b,i_1,i_2,...,i_{n-1}]}\,.\nn
\end{eqnarray}
Thus we can start with the set of $n$-forms with polynomial coefficient
functions and then use the polynomial reduction procedures to obtain 
independent $n$-forms on a smooth algebraic variety. The reduction of exact
forms can be done in a similar way, where one first generates exact forms and
then interprets them as polynomials in the auxiliary coordinate ring. The
reductions steps work as described above.


\begin{thebibliography}{99}

\bibitem{2gamNNLO}
  S.~Catani, L.~Cieri, D.~de Florian, G.~Ferrera and M.~Grazzini,
  ``Diphoton production at hadron colliders: a fully-differential QCD calculation at NNLO,''
  Phys.\ Rev.\ Lett.\  {\bf 108} (2012) 072001
  [arXiv:1110.2375 [hep-ph]].

\bibitem{TopNNLO}
  M.~Czakon, P.~Fiedler and A.~Mitov,
  ``Resolving the Tevatron Top Quark Forward-Backward Asymmetry Puzzle: Fully Differential Next-to-Next-to-Leading-Order Calculation,''
  Phys.\ Rev.\ Lett.\  {\bf 115} (2015) 5,  052001
  [arXiv:1411.3007 [hep-ph]].

\bibitem{VVNNLO}
  T.~Gehrmann, M.~Grazzini, S.~Kallweit, P.~Maierhöfer, A.~von Manteuffel, S.~Pozzorini, D.~Rathlev and L.~Tancredi,
  ``$W^+W^-$ Production at Hadron Colliders in Next to Next to Leading Order QCD,''
  Phys.\ Rev.\ Lett.\  {\bf 113} (2014) 21,  212001
  [arXiv:1408.5243 [hep-ph]];
%
  F.~Cascioli {\it et al.},
  ``ZZ production at hadron colliders in NNLO QCD,''
  Phys.\ Lett.\ B {\bf 735} (2014) 311
  [arXiv:1405.2219 [hep-ph]].

\bibitem{ZgamNNLO}
  M.~Grazzini, S.~Kallweit, D.~Rathlev and A.~Torre,
  ``$Z\gamma$ production at hadron colliders in NNLO QCD,''
  Phys.\ Lett.\ B {\bf 731} (2014) 204
  [arXiv:1309.7000 [hep-ph]].

\bibitem{VjetNNLO}
  R.~Boughezal, C.~Focke, X.~Liu and F.~Petriello,
  ``$W$-boson production in association with a jet at next-to-next-to-leading order in perturbative QCD,''
  Phys.\ Rev.\ Lett.\  {\bf 115} (2015) 6,  062002
  [arXiv:1504.02131 [hep-ph]];
%
  A.~Gehrmann-De Ridder, T.~Gehrmann, E.~W.~N.~Glover, A.~Huss and T.~A.~Morgan,
  ``Precise QCD predictions for the production of a Z boson in association with a hadronic jet,''
  arXiv:1507.02850 [hep-ph].

\bibitem{HjetNNLO}
   X.~Chen, T.~Gehrmann, E.~W.~N.~Glover and M.~Jaquier,
   ``Precise QCD predictions for the production of Higgs + jet final states,''
   Phys.\ Lett.\ B {\bf 740} (2015) 147
   [arXiv:1408.5325 [hep-ph]];
%
   R.~Boughezal, C.~Focke, W.~Giele, X.~Liu and F.~Petriello,
   ``Higgs boson production in association with a jet at NNLO using jettiness subtraction,''
   Phys.\ Lett.\ B {\bf 748} (2015) 5
   [arXiv:1505.03893 [hep-ph]];
%
   R.~Boughezal, F.~Caola, K.~Melnikov, F.~Petriello and M.~Schulze,
   ``Higgs boson production in association with a jet at next-to-next-to-leading order,''
   Phys.\ Rev.\ Lett.\  {\bf 115} (2015) 8,  082003
   [arXiv:1504.07922 [hep-ph]].

\bibitem{HN3LO}
  C.~Anastasiou, C.~Duhr, F.~Dulat, F.~Herzog and B.~Mistlberger,
  ``Higgs Boson Gluon-Fusion Production in QCD at Three Loops,''
  Phys.\ Rev.\ Lett.\  {\bf 114} (2015) 212001
  [arXiv:1503.06056 [hep-ph]].




\bibitem{VBFNLO}
 K.~Arnold {\it et al.},
 ``VBFNLO: A Parton level Monte Carlo for processes with electroweak bosons,''
 Comput.\ Phys.\ Commun.\  {\bf 180} (2009) 1661
 [arXiv:0811.4559 [hep-ph]].

\bibitem{BlackHat}
 C.~F.~Berger, Z.~Bern, L.~J.~Dixon, F.~Febres Cordero, D.~Forde, H.~Ita, D.~A.~Kosower and D.~Maitre,
 ``An Automated Implementation of On-Shell Methods for One-Loop Amplitudes,''
 Phys.\ Rev.\ D {\bf 78} (2008) 036003
 [arXiv:0803.4180 [hep-ph]].

\bibitem{NJet}
 S.~Badger, B.~Biedermann and P.~Uwer,
 ``NGluon: A Package to Calculate One-loop Multi-gluon Amplitudes,''
 Comput.\ Phys.\ Commun.\  {\bf 182} (2011) 1674
 [arXiv:1011.2900 [hep-ph]].

\bibitem{MadLoop}
 V.~Hirschi, R.~Frederix, S.~Frixione, M.~V.~Garzelli, F.~Maltoni and R.~Pittau,
 ``Automation of one-loop QCD corrections,''
 JHEP {\bf 1105} (2011) 044
 [arXiv:1103.0621 [hep-ph]].

\bibitem{HelacNLO}
 G.~Bevilacqua, M.~Czakon, M.~V.~Garzelli, A.~van Hameren, A.~Kardos, C.~G.~Papadopoulos, R.~Pittau and M.~Worek,
 ``Helac-nlo,''
 Comput.\ Phys.\ Commun.\  {\bf 184} (2013) 986
 [arXiv:1110.1499 [hep-ph]].

\bibitem{GoSam}
 G.~Cullen, N.~Greiner, G.~Heinrich, G.~Luisoni, P.~Mastrolia, G.~Ossola, T.~Reiter and F.~Tramontano,
 ``Automated One-Loop Calculations with GoSam,''
 Eur.\ Phys.\ J.\ C {\bf 72} (2012) 1889
 [arXiv:1111.2034 [hep-ph]];
 G.~Cullen {\it et al.},
 ``G$\scriptsize{O}$S$\scriptsize{AM}$-2.0: a tool for automated one-loop calculations within the Standard Model and beyond,''
 Eur.\ Phys.\ J.\ C {\bf 74} (2014) 8,  3001
 [arXiv:1404.7096 [hep-ph]].

\bibitem{OpenLoops}
 F.~Cascioli, P.~Maierhofer and S.~Pozzorini,
 ``Scattering Amplitudes with Open Loops,''
 Phys.\ Rev.\ Lett.\  {\bf 108} (2012) 111601
 [arXiv:1111.5206 [hep-ph]].

\bibitem{Recola}
 S.~Actis, A.~Denner, L.~Hofer, A.~Scharf and S.~Uccirati,
 ``Recursive generation of one-loop amplitudes in the Standard Model,''
 JHEP {\bf 1304} (2013) 037
 [arXiv:1211.6316 [hep-ph]].

\bibitem{DDReduction}
 A.~Denner and S.~Dittmaier,
 ``Reduction schemes for one-loop tensor integrals,''
 Nucl.\ Phys.\ B {\bf 734} (2006) 62
 [hep-ph/0509141].


\bibitem{UnitarityI}
 Z.~Bern, L.~J.~Dixon, D.~C.~Dunbar and D.~A.~Kosower,
 ``One loop n point gauge theory amplitudes, unitarity and collinear limits,''
 Nucl.\ Phys.\ B {\bf 425} (1994) 217
 [hep-ph/9403226].

\bibitem{GenUnitarityI}
 Z.~Bern, L.~J.~Dixon, D.~C.~Dunbar and D.~A.~Kosower,
 ``Fusing gauge theory tree amplitudes into loop amplitudes,''
 Nucl.\ Phys.\ B {\bf 435} (1995) 59
 [hep-ph/9409265].

\bibitem{GenUnitarityII}
 R.~Britto, F.~Cachazo and B.~Feng,
 ``Generalized unitarity and one-loop amplitudes in N=4 super-Yang-Mills,''
 Nucl.\ Phys.\ B {\bf 725} (2005) 275
 [hep-th/0412103].

\bibitem{GenUnitarityIII}
 A.~Brandhuber, S.~McNamara, B.~J.~Spence and G.~Travaglini,
 ``Loop amplitudes in pure Yang-Mills from generalised unitarity,''
 JHEP {\bf 0510} (2005) 011
 [hep-th/0506068];
%
 S.~D.~Badger,
 ``Direct Extraction Of One Loop Rational Terms,''
 JHEP {\bf 0901} (2009) 049
 [arXiv:0806.4600 [hep-ph]].
%
 D.~Forde,
 ``Direct extraction of one-loop integral coefficients,''
 Phys.\ Rev.\ D {\bf 75} (2007) 125019
 [arXiv:0704.1835 [hep-ph]].

\bibitem{OPP}
 G.~Ossola, C.~G.~Papadopoulos and R.~Pittau,
 ``Reducing full one-loop amplitudes to scalar integrals at the integrand level,''
 Nucl.\ Phys.\ B {\bf 763} (2007) 147
 [hep-ph/0609007].

\bibitem{NumUnitarity}
 R.~K.~Ellis, W.~T.~Giele and Z.~Kunszt,
 ``A Numerical Unitarity Formalism for Evaluating One-Loop Amplitudes,''
 JHEP {\bf 0803} (2008) 003
 [arXiv:0708.2398 [hep-ph]];
 W.~T.~Giele, Z.~Kunszt and K.~Melnikov,
 ``Full one-loop amplitudes from tree amplitudes,''
 JHEP {\bf 0804} (2008) 049
 [arXiv:0801.2237 [hep-ph]].
 R.~K.~Ellis, W.~T.~Giele, Z.~Kunszt and K.~Melnikov,
 ``Masses, fermions and generalized $D$-dimensional unitarity,''
 Nucl.\ Phys.\ B {\bf 822} (2009) 270
 [arXiv:0806.3467 [hep-ph]];

\bibitem{IBP}
 K.~G.~Chetyrkin and F.~V.~Tkachov,
 ``Integration by Parts: The Algorithm to Calculate beta Functions in 4 Loops,''
 Nucl.\ Phys.\ B {\bf 192} (1981) 159;

\bibitem{Reduce}
 A.~von Manteuffel and C.~Studerus,
 ``Reduze 2 - Distributed Feynman Integral Reduction,''
 arXiv:1201.4330 [hep-ph].
    %
 C.~Studerus,
 ``Reduze-Feynman Integral Reduction in C++,''
 Comput.\ Phys.\ Commun.\  {\bf 181} (2010) 1293
 [arXiv:0912.2546 [physics.comp-ph]].
 A.~von Manteuffel and R.~M.~Schabinger,
 ``A novel approach to integration by parts reduction,''
 Phys.\ Lett.\ B {\bf 744} (2015) 101
 [arXiv:1406.4513 [hep-ph]].

\bibitem{Air}
 C.~Anastasiou and A.~Lazopoulos,
 ``Automatic integral reduction for higher order perturbative calculations,''
 JHEP {\bf 0407} (2004) 046
 [hep-ph/0404258].

\bibitem{Fire}
 A.~V.~Smirnov and V.~A.~Smirnov,
 ``FIRE4, LiteRed and accompanying tools to solve integration by parts relations,''
 Comput.\ Phys.\ Commun.\  {\bf 184} (2013) 2820
 [arXiv:1302.5885 [hep-ph]];
 A.~V.~Smirnov and V.~A.~Smirnov,
 ``FIRE4, LiteRed and accompanying tools to solve integration by parts relations,''
 Comput.\ Phys.\ Commun.\  {\bf 184} (2013) 2820
 [arXiv:1302.5885 [hep-ph]];
\bibitem{Smirnov:2014hma}
  A.~V.~Smirnov,
  ``FIRE5: a C++ implementation of Feynman Integral REduction,''
  Comput.\ Phys.\ Commun.\  {\bf 189} (2014) 182
  [arXiv:1408.2372 [hep-ph]].


\bibitem{LiteRed}
  R.~N.~Lee,
  ``LiteRed 1.4: a powerful tool for reduction of multiloop integrals,''
  J.\ Phys.\ Conf.\ Ser.\  {\bf 523} (2014) 012059
  [arXiv:1310.1145 [hep-ph]].

\bibitem{Laporta}
 S.~Laporta,
 ``High precision calculation of multiloop Feynman integrals by difference equations,''
 Int.\ J.\ Mod.\ Phys.\ A {\bf 15} (2000) 5087
 [hep-ph/0102033].

\bibitem{MZProgram}
 R.~K.~Ellis, W.~T.~Giele, Z.~Kunszt, K.~Melnikov and G.~Zanderighi,
 JHEP {\bf 0901} (2009) 012
 [arXiv:0810.2762 [hep-ph]].


\bibitem{W3jB}
 C.~F.~Berger {\it et al.},
 ``Precise Predictions for $W$ + 3 Jet Production at Hadron Colliders,''
 Phys.\ Rev.\ Lett.\  {\bf 102} (2009) 222001
 [arXiv:0902.2760 [hep-ph]].

\bibitem{W3jR}
 R.~K.~Ellis, K.~Melnikov and G.~Zanderighi,
 ``W+3 jet production at the Tevatron,''
 Phys.\ Rev.\ D {\bf 80} (2009) 094002
 [arXiv:0906.1445 [hep-ph]].

\bibitem{W4j}
 C.~F.~Berger {\it et al.},
 ``Precise Predictions for W + 4 Jet Production at the Large Hadron Collider,''
 Phys.\ Rev.\ Lett.\  {\bf 106} (2011) 092001
 [arXiv:1009.2338 [hep-ph]].

\bibitem{W5j}
 Z.~Bern, L.~J.~Dixon, F.~Febres Cordero, S.~Höche, H.~Ita, D.~A.~Kosower, D.~Maître and K.~J.~Ozeren,
 ``Next-to-Leading Order $W + 5$-Jet Production at the LHC,''
 Phys.\ Rev.\ D {\bf 88} (2013) 1,  014025
 [arXiv:1304.1253 [hep-ph]].

\bibitem{5Jets}
 S.~Badger, B.~Biedermann, P.~Uwer and V.~Yundin,
 ``Next-to-leading order QCD corrections to five jet production at the LHC,''
 Phys.\ Rev.\ D {\bf 89} (2014) 3,  034019
 [arXiv:1309.6585 [hep-ph]].

\bibitem{2LoopUnitarityQCD}
 Z.~Bern, L.~J.~Dixon and D.~A.~Kosower,
 ``A Two loop four gluon helicity amplitude in QCD,''
 JHEP {\bf 0001} (2000) 027
 [hep-ph/0001001].
%
 Z.~Bern, A.~De Freitas and L.~J.~Dixon,
 ``Two loop helicity amplitudes for gluon-gluon scattering in QCD and supersymmetric Yang-Mills theory,''
 JHEP {\bf 0203} (2002) 018
 [hep-ph/0201161].
%
 Z.~Bern, A.~De Freitas and L.~J.~Dixon,
 ``Two loop helicity amplitudes for quark gluon scattering in QCD and gluino gluon scattering in supersymmetric Yang-Mills theory,''
 JHEP {\bf 0306} (2003) 028
 [JHEP {\bf 1404} (2014) 112]
 [hep-ph/0304168].

\bibitem{MultiLoopUnitarity}
 Z.~Bern, L.~J.~Dixon, D.~C.~Dunbar, M.~Perelstein and J.~S.~Rozowsky,
 ``On the relationship between Yang-Mills theory and gravity and its implication for ultraviolet divergences,''
 Nucl.\ Phys.\ B {\bf 530}, 401 (1998)
 [hep-th/9802162];


\bibitem{UVGravity}
 Z.~Bern, J.~J.~Carrasco, L.~J.~Dixon, H.~Johansson and R.~Roiban,
 ``The Ultraviolet Behavior of N=8 Supergravity at Four Loops,''
 Phys.\ Rev.\ Lett.\  {\bf 103} (2009) 081301
 [arXiv:0905.2326 [hep-th]].

\bibitem{BDS}
  Z.~Bern, L.~J.~Dixon and V.~A.~Smirnov,
  ``Iteration of planar amplitudes in maximally supersymmetric Yang-Mills theory at three loops and beyond,''
  Phys.\ Rev.\ D {\bf 72} (2005) 085001
  [hep-th/0505205].
 
\bibitem{SYMUnitarity}
  N.~Arkani-Hamed, J.~L.~Bourjaily, F.~Cachazo, S.~Caron-Huot and J.~Trnka,
  ``The All-Loop Integrand For Scattering Amplitudes in Planar N=4 SYM,''
  JHEP {\bf 1101} (2011) 041
  [arXiv:1008.2958 [hep-th]];
    %
  J.~L.~Bourjaily and J.~Trnka,
  ``Local Integrand Representations of All Two-Loop Amplitudes in Planar SYM,''
  JHEP {\bf 1508} (2015) 119
  [arXiv:1505.05886 [hep-th]].

\bibitem{2LoopAmpl5g}
 S.~Badger, H.~Frellesvig and Y.~Zhang,
 ``A Two-Loop Five-Gluon Helicity Amplitude in QCD,''
 JHEP {\bf 1312} (2013) 045
 [arXiv:1310.1051 [hep-ph]];
 S.~Badger, G.~Mogull, A.~Ochirov and D.~O'Connell,
 ``A Complete Two-Loop, Five-Gluon Helicity Amplitude in Yang-Mills Theory,''
 arXiv:1507.08797 [hep-ph].

\bibitem{MLoopParamMO}
 P.~Mastrolia and G.~Ossola,
 ``On the Integrand-Reduction Method for Two-Loop Scattering Amplitudes,''
 JHEP {\bf 1111} (2011) 014
 [arXiv:1107.6041 [hep-ph]].

\bibitem{MLoopParamBFZ}
 S.~Badger, H.~Frellesvig and Y.~Zhang,
 ``Hepta-Cuts of Two-Loop Scattering Amplitudes,''
 JHEP {\bf 1204} (2012) 055
 [arXiv:1202.2019 [hep-ph]].

\bibitem{MLoopAlgGeo}
 Y.~Zhang,
 ``Integrand-Level Reduction of Loop Amplitudes by Computational Algebraic Geometry Methods,''
 JHEP {\bf 1209} (2012) 042
 [arXiv:1205.5707 [hep-ph]].

\bibitem{2LoopMaxCuts}
 E.~I.~Buchbinder and F.~Cachazo,
 ``Two-loop amplitudes of gluons and octa-cuts in N=4 super Yang-Mills,''
 JHEP {\bf 0511} (2005) 036
 [hep-th/0506126];
 F.~Cachazo,
 ``Sharpening The Leading Singularity,''
 arXiv:0803.1988 [hep-th];
 Z.~Bern, J.~J.~M.~Carrasco, H.~Johansson and D.~A.~Kosower,
 ``Maximally supersymmetric planar Yang-Mills amplitudes at five loops,''
 Phys.\ Rev.\ D {\bf 76} (2007) 125020
 [arXiv:0705.1864 [hep-th]].

\bibitem{MLoopContours}
 D.~A.~Kosower and K.~J.~Larsen,
 ``Maximal Unitarity at Two Loops,''
 Phys.\ Rev.\ D {\bf 85} (2012) 045017
 [arXiv:1108.1180 [hep-th]];
%

\bibitem{MLoopContoursCycles}
 S.~Caron-Huot and K.~J.~Larsen,
 ``Uniqueness of two-loop master contours,''
 JHEP {\bf 1210} (2012) 026
 [arXiv:1205.0801 [hep-ph]].

\bibitem{MLoopCohom}
 A.~Georgoudis and Y.~Zhang,
 ``Two-loop Integral Reduction from Elliptic and Hyperelliptic Curves,''
 arXiv:1507.06310 [hep-th].

\bibitem{KosowerIBP} 
 J.~Gluza, K.~Kajda and D.~A.~Kosower,
 ``Towards a Basis for Planar Two-Loop Integrals,''
 Phys.\ Rev.\ D {\bf 83}, 045012 (2011)
 [arXiv:1009.0472 [hep-th]].

\bibitem{sIBP}
 R.~M.~Schabinger,
 ``A New Algorithm For The Generation Of Unitarity-Compatible Integration By Parts Relations,''
 JHEP {\bf 1201} (2012) 077
 [arXiv:1111.4220 [hep-ph]].
         
\bibitem{IBPDiffGeom}
 Y.~Zhang,
 ``Integration-by-parts identities from the viewpoint of differential geometry,''
 arXiv:1408.4004 [hep-th].

\bibitem{YZhangElliptic} 
 M.~S{\o}gaard and Y.~Zhang,
 ``Elliptic Functions and Maximal Unitarity,''
 Phys.\ Rev.\ D {\bf 91}, no. 8, 081701 (2015)
 [arXiv:1412.5577 [hep-th]].

\bibitem{YZhangElliptic2} 
 A.~Georgoudis and Y.~Zhang,
 ``Two-loop Integral Reduction from Elliptic and Hyperelliptic Curves,''
 arXiv:1507.06310 [hep-th].

\bibitem{NVbasis}
 W.~L.~van Neerven and J.~A.~M.~Vermaseren,
 ``Large Loop Integrals,''
 Phys.\ Lett.\ B {\bf 137} (1984) 241.

\bibitem{IntegralsExplicit}
  G.~'t Hooft and M.~J.~G.~Veltman,
  ``Scalar One Loop Integrals,''
  Nucl.\ Phys.\  B {\bf 153}, 365 (1979);
  %
  G.~J.~van Oldenborgh and J.~A.~M.~Vermaseren,
  ``New Algorithms for One Loop Integrals,''
  Z.\ Phys.\  C {\bf 46}, 425 (1990);
  %
  W.~Beenakker and A.~Denner,
  ``Infrared Divergent Scalar Box Integrals With Applications in the
  Electroweak Standard Model,''
  Nucl.\ Phys.\  B {\bf 338}, 349 (1990);
  %
  A.~Denner, U.~Nierste and R.~Scharf,
  ``A Compact expression for the scalar one loop four point function,''
  Nucl.\ Phys.\  B {\bf 367}, 637 (1991);
  %
  Z.~Bern, L.~J.~Dixon and D.~A.~Kosower,
  ``Dimensionally regulated pentagon integrals,''
  Nucl.\ Phys.\  B {\bf 412} (1994) 751
  [hep-ph/9306240];
  %
  T.~Hahn and M.~P\'erez-Victoria,
  ``Automatized one-loop calculations in four and D dimensions,''
  Comput.\ Phys.\ Commun.\  {\bf 118}, 153 (1999)
  [hep-ph/9807565];
  %
  R.~K.~Ellis and G.~Zanderighi,
  ``Scalar one-loop integrals for QCD,''
  JHEP {\bf 0802}, 002 (2008)
  [0712.1851 [hep-ph]].

\bibitem{RevUnitarity}
 H.~Ita,
 ``Susy Theories and QCD: Numerical Approaches,''
 J.\ Phys.\ A {\bf 44} (2011) 454005
 [arXiv:1109.6527 [hep-th]].


\bibitem{Dirschmid}
  H.J.~Dirschmid,
    ``Tensoren und Felder,'', Springer-Verlag/Vienna (1996);
    %
  M.~Nakahara,
    ``Geometry, topology and physics,''
      Boca Raton, USA: Taylor \& Francis (2003) 573 p

\bibitem{Tarasov97}
 O.~V.~Tarasov,
 ``Generalized recurrence relations for two loop propagator integrals with arbitrary masses,''
 Nucl.\ Phys.\ B {\bf 502} (1997) 455
 [hep-ph/9703319].

\bibitem{SchabingerRadcor13}
 R.~M.~Schabinger (presenter),  A.~von~Manteuffel, 
 "The Two-Loop Analog of the Passarino-Veltman Result And Beyond",
 PoS RADCOR {\bf 2013} (2013).

\bibitem{Mathematica}
 S. Wolfram, {\it The Mathematica book}, 5th edition, Wolfram Media, Inc., 2003.

\bibitem{NumeratorsFH}
  B.~Feng and R.~Huang,
  ``The classification of two-loop integrand basis in pure four-dimension,''
  JHEP {\bf 1302} (2013) 117
  [arXiv:1209.3747 [hep-ph]].

\bibitem{Hartshorne}
  R.Hartshorne,
    ``Algebraic Geometry,'', Springer-Verlag/New York (1977)
    %

\bibitem{DOlive}
R.~J.~Eden, P.~V.~Landshoff, D.~I.~Olive, J.~C.~Polkinghorne,
  ``The Analytic S-Matrix,'' Cambridge University Press (1966) 


\end{thebibliography}
\end{document}